\documentclass[onecolumn,aps,groupedaddress,nobibnotes,nofootinbib]{revtex4-2} 

\usepackage{graphicx}  
\graphicspath{{../pdf/}{C:/Users/Figures}}
\usepackage{dcolumn}   
\usepackage{bm}        
\usepackage{amssymb}   
\usepackage{amsmath}   
\usepackage{braket}    
\usepackage[caption=false]{subfig}
\usepackage{hyperref}
\usepackage{color}
\usepackage{ulem}
\newcommand{\be}[1]{\begin{equation}\label{#1}}
\newcommand{\ee}{\end{equation}}
\newcommand{\bea}[1]{\begin{eqnarray}\label{#1}}
\newcommand{\eea}{\end{eqnarray}}
\newcommand{\no}{\nonumber \\}
\newcommand{\Fig}[1]{Fig.(\ref{#1})}
\newcommand{\Eq}[1]{Eq.(\ref{#1})}
\newcommand{\Sec}[1]{Section~\ref{#1}}
\newcommand{\bsub}{\begin{subequations}}
\newcommand{\esub}{\end{subequations}}
\newcommand{\bwt}{\begin{widetext}}
\newcommand{\ewt}{\end{widetext}}
\def\Fig#1{Fig.(\ref{#1})}

\def\myoverDefn#1#2{\hbox{\space \raise-2mm\hbox{$\textstyle{#1} \atop \scriptstyle{#2}$} }}
\def\defn{\overset{\trm{def}}{=}}

\def\dag{\dagger}

\def\s{\sigma}

\def\G{\Gamma}

\def\IP#1#2{\langle #1| #2 \rangle}              
\def\EV#1#2#3{\langle #1| #2 | #3\rangle}   
\def\dfrac#1#2{\displaystyle{\frac{#1}{#2}}} 
\def\M{\mathcal{M}}

\def\S{\Sigma}
\def\ket#1{|#1\rangle}
\def\bra#1{\langle #1|}

\usepackage{mathtools,amsthm}
\DeclareMathOperator{\Tr}{Tr}

\def\a{\alpha}

\def\E{\mathcal{E}}
\def\M{\mathcal{M}}
\def\MN{\mathcal{M}^{N}}
\def\F{\mathcal{F}}
\def\HS{\mathcal{HS}}
\def\SN{\mathcal{S}_N}
\def\SH{\mathcal{S}(\mathcal{H})}

\def\re#1{\textrm{Re}\left({#1}\right)}

\def\R{\mathbb{R}}
\def\C{\mathbb{C}}
\def\H{\mathcal{H}}

\def\s{\sigma}

\def\thalf{\tfrac{1}{2}}
\def\half{\frac{1}{2}}

\def\bZ{\bar{Z}}

\def\thalf{\tfrac{1}{2}\,}
\def\dhalf{\dfrac{1}{2}\,}
\def\pd{\partial}

\def\bfB{{\mathbf{B}}}

\def\bfx{{\mathbf{x}}}
\def\bfxhat{{\hat{\mathbf{x}}}}
\def\bfxdot{{\dot{\mathbf{x}}}}
\def\bfzhat{{\hat{\mathbf{z}}}}
\def\bfadot{\dot{\mathbf{a}}}
\def\bftau{{\boldsymbol{\tau}}}

\def\bfy{{\mathbf{y}}}
\def\bfg{{\mathbf{g}}}
\def\bfa{{\mathbf{a}}}

\def\tit#1{\textit{#1}}
\def\trm#1{\textrm{#1}}
\def\bfsigma{\boldsymbol{\sigma}}

\def\TE{T(\E)}

\def\HE{H(\E)}
\def\VE{V(\E)}

\def\inE{\in\E}

\def\tauopr{\rho_1^{1/2}\,\rho_2\,\rho_1^{1/2}}
\def\roottauopr{\sqrt{\rho_1^{1/2}\,\rho_2\,\rho_1^{1/2}}}

\def\rootF{\sqrt{F}(\rho_1,\rho_2)}
\def\tr#1{\textrm{Tr}\left[#1\right]}
\def\HS{\mathcal{HS}}

\def\MN{\mathcal{M}^{(N)}}
\def\D{\mathcal{D}}
\def\P{\mathcal{P}}
\def\dA{\dot{A}}
\def\T{\mathcal{T}}
\def\N{\mathcal{N}}
\def\Id{\mathbb{I}}
\def\l{\lambda}
\def\S{\mathbb{S}}

\def\rootrhoone{\sqrt{\rho_1}}
\def\rootrhotwo{\sqrt{\rho_2}}
\def\rhoonehalf{\rho_1^{1/2}}

\def\rootrhoonehalf{\rho_1^{1/2}}
\def\rootrhotwohalf{\rho_2^{1/2}}
\def\rhooneinvhalf{\rho_1^{-1/2}}
\def\rhotwoinvhalf{\rho_2^{-1/2}}
\def\roottau{\sqrt{\tau}}
\def\Ainv{A^{-1}}
\def\Adaginv{A^{-1\dag}}
\def\dX{\dot{X}}
\def\L{\mathcal{L}}
\def\dZ{\dot{Z}}
\def\Zbar{\bar{Z}}
\def\dZbar{\dot{\bar{Z}}}
\def\sstar{s^{*}}
\def\rhostar{\rho_{s^*}}
\def\Astar{A_{s^*}}
\def\Mstar{M_{s^*}}
\def\cosstar{\cos(s^*)}
\def\sinstar{\sin(s^*)}
\def\tanstar{\tan(s^*)}
\def\Mstaropr{\rho_0^{-1/2} \,\sqrt{\rho_0^{1/2} \rhostar \rho_0^{1/2}} \, \rho_0^{-1/2} }
\def\cossqrdstar{\cos^2(s^*)}
\def\sinsqrdstar{\sin^2(s^*)}

\def\roottauoprstar{ \sqrt{\rho_0^{1/2} \rho_{s^*} \rho_0^{1/2} }}
\def\rootFstar{\sqrt{F_{s^*}}}
\def\rootoneminusFstar{\sqrt{1-F_{s^*}}}

\def\Id{\mathbb{I}}

\def\GHZ{\ket{GHZ}}
\def\W{\ket{W}}
\def\rhoGHZpure{\ket{GHZ}\bra{GHZ}}
\def\rhoWpure{\ket{W}\bra{W}}

\begin{document}

\title{Geodesics for  mixed quantum states via their geometric mean operator}

\author{Paul M. Alsing$^{*}$, Carlo Cafaro$^{*\,\dag}$}
\affiliation{$^{*}$University at Albany-SUNY, Albany, NY 12222, USA}
\email{corresponding author: palsing@albany.edu; alsingpm@gmail.com}
\affiliation{$^{\dag}$SUNY Polytechnic Institute, Utica, NY 13502, USA}
\author{Shannon Ray$^{\ddag}$}
\affiliation{$^{\ddag}$Griffiss Institute, Rome, NY, 13411, USA}


\date{\today}

\begin{abstract}
In this work we examine the geodesic between two mixed states of arbitrary dimension by means of their geometric mean operator. 
We utilize the fiber bundle approach by which the distance between two mixed state density operators $\rho_1$ and $\rho_2$ in the base space $\M$ is given by the shortest distance in the (Hilbert Schmidt) bundle space $\E$ of their purifications. The latter is well-known to be given by the Bures distance along the horizontal lift in $\E$ of the  geodesic between the $\rho_1$ and $\rho_2$ in $\M$. The horizontal lift is that unique curve in $\E$ that orthogonally traverses the fibers $\F\subset\E$ above the curve in $\M$, and projects down onto it.  
We briefly review this formalism and show how it can be used to construct the intermediate mixed quantum states $\rho(s)$ along the base space geodesic parameterized by affine parameter $s$ between the initial $\rho_1$ and final $\rho_2$ states. We emphasize the role played by geometric mean operator 
$M(s) = \rho_1^{-1/2}\, \sqrt{\rho_1^{1/2}\rho(s)\rho_1^{1/2}}\,\rho_1^{-1/2}$, 
where the Uhlmann root fidelity between $\rho_1$ and   $\rho(s)$ is given by
$\sqrt{F}(\rho_1,\rho(s)) = \Tr[M(s)\,\rho_1] = \Tr[\sqrt{\rho_1^{1/2}\rho(s)\rho_1^{1/2}}]$, and
$\rho(s) = M(s)\,\rho_1\,M(s)$. We give  examples for the geodesic between the maximally mixed state and a pure state in arbitrary dimensions, as well as for the geodesic
between Werner states
$\rho(p) = (1-p) \Id/N + p\,\ket{\Psi}\bra{\Psi}$ 
with 
$\ket{\Psi} = \{\ket{GHZ}, \ket{W}\}$
in dimension $N=2^3$. For the latter, we compare expressions in the limit $p\to1$ to the infinite number of possible geodesics between the orthogonal pure states $\ket{GHZ}$ and $\ket{W}$.
Lastly, we compute the analytic form for the  density matrices along the geodesic that connects two arbitrary endpoint qubit density matrices within the Bloch ball for dimension $N=2$.
\end{abstract}

\maketitle

\section{\label{sec:intro} Introduction}
The role of the geometry of quantum states \cite{Zyczkowski_2ndEd:2020} has played an ever increasing role in diverse areas of physics and  quantum information including metrology \cite{Sjoqvist:2000},  atomic and molecular physics \cite{Mead:1992},  phase transitions \cite{Carollo:2020},   and even quantum computing \cite{Nielsen:2006} (for a recent review see \cite{Chien:2023} and references therein). 
%
It is known that if the distance between density matrices for both pure and
mixed states expresses statistical distinguishability, then this distance
must decrease under randomization (i.e., coarse-graining)
\cite{petz96a,petz99}. In the case of pure states for closed quantum systems, the
Fubini-Study distance is the only natural choice for a measure that defines
\textquotedblleft random states\textquotedblright\ since the Fubini-Study
metric is the only monotone Riemannian metric. However, practical scenarios
occur mainly for open system dynamics in mixed quantum states 
\cite{taddei13,adolfo13,deffner13}. 
\smallskip

Unlike the geometry of pure states,
geometrical investigations of mixed states exhibit several additional
delicate issues. Firstly, there is the possibility of the emergence of
arbitrarily complicated nonunitary quantum evolutions specified by a master
equation in the Lindblad form. As a consequence, it can become pretty
difficult to carry out analytical calculations relative to the effective
dynamical trajectories produced by an open quantum system in a mixed quantum
state \cite{carlini08,brody19}. For an insightful presentation on conceptual
and computational challenges in obtaining the time-optimal quantum evolution
of mixed states ruled by a master equation, we suggest Ref. \cite{carlini08}.
\smallskip

In reality, even focusing on unitary evolutions of closed systems, we can
find tangible differences in the geometry of quantum evolutions of systems
in pure and mixed states. For instance, optimal-time evolutions of pure
(mixed) states are generally specified by constant (time-varying)
Hamiltonian operators \cite{hornedal22}. Secondly, as stated in the
Morozova-Cencov-Petz theorem \cite{petz96a,petz99}, there exist infinitely
many monotone Riemannian metrics on the space of mixed states. Perhaps, the
chief obstacle when geometrically studying the dynamics of open systems in
mixed quantum states is exactly the nonuniqueness of the metric. Indeed, the
freedom to select a variety of distance measures between mixed states, leads
one to explore the use of differently motivated measures, each one with their own
particular set of strengths and weaknesses. 
\smallskip

However, in general, it is usually difficult
to find exact analytical expressions of geodesic paths on general manifolds
of mixed quantum states furnished with statistically meaningful Riemann
metrics, even under the assumption of having selected a metric and having at
hand the effective dynamical trajectory produced by the system. This task
can be carried out with success in a handful of scenarios \cite{weis13}. For
instance, geodesic paths that join two mixed states are known for the
Quantum Fisher Information (QFI) metric \cite{uhlmann95}, the Wigner-Yanase
metric \cite{gibilisco03} and, finally, the metric originating from the
trace distance \cite{cai17}. Recalling that the link between the QFI metric
and the Bures metric \cite{bures69,uhlmann76,hubner92} is $g_{\mu \nu
}^{\left( \mathrm{QFI}\right) }=4g_{\mu \nu }^{\left( \mathrm{Bures}\right) }$, 
expressions for geodesic paths can be recast by means of projections of
large circles on a sphere in a purifying space in this case \cite{dittman95}. 
Moreover, explicit formulas for the geodesic path, geodesic distance, and,
finally, sectional and scalar curvatures can be obtained in the
Wigner-Yanase metric case \cite{wigner63,luo03}. Extending the classical
pull-back approach to the Fisher information to the quantum setting, these
expressions were first presented by Gibilisco and Isola in Ref. \cite%
{gibilisco03}. For a nice review on the geometry of quantum state spaces,
both pure and mixed, we indicate Ref. \cite{uhlmann09}.
\bigskip

Much of the interest in the geometry of pure and mixed quantum states was inspired and 
spurred on by investigations of Berry's geometric phase and its applicability  to a wide range of practical physical problems (for extensive reviews see  \cite{Wilczek:1989,Bohm:2003,Jamiolkowski:2004}). Simon \cite{Simon:1983} and subsequent authors introduced the concept of the geometric phase as the holonomy of a path traced out by a pure state $\ket{\Psi}$ (in an $N$ dimensional Hilbert space) in the  higher dimensional (bundle) space $\E$ as it traverses horizontally across the $U(N)$ fibers $\F$ that live above each point of the projected closed path in the base space $\M$ of projective Hilbert space 
$\mathbb{C}P^{N-1}$. The division of the tangent space of $\E$ into a (direct) sum of vertical vectors (along the fibers) and orthogonal horizontal vectors is given by the introduction of a connection, whose properties determine the definition of the horizontal subspace. A path that traverses the fibers $\F$ of $\E$ with tangent vectors lying solely in the horizontal subspace is said to be parallel transported. The total relative phase of the pure state in $\E$ when it returns to the same fiber (but not necessarily the same point in the fiber)  has two contributions: a dynamic phase that depends on the Hamiltonian that evolves the state, and a geometric phase that depends solely on the horizontal lift in $\E$ of the closed circuit path in $\M$. The geometric phase is given by the closed integral of the (exterior) derivative of the connection $\mathcal{A} = \trm{Im}[\IP{\Psi}{\dot{\Psi}}]$  along the path, or equivalently by Stoke's theorem, the integral of the (gauge invariant) curvature $\F_{\mathcal{A}}=d\mathcal{A}$ over a surface bounded by the closed path. In addition, the horizontal lift is a geodesic, whose length determines the distance between the two pure states, which plays a prominent role in constructing measures of distinguishability between quantum states.
\bigskip

Over the past 35 years the above ideas applicable to pure states $\ket{\Psi}$ have been extensively explored and extended to mixed states $\rho$ in $\M$, with seminal early contributions from  authors such as Uhlmann \cite{Uhlmann:1986,Uhlmann:1991,Uhlmann:O3orbits:1993,Uhlmann:DenOprs:DiffGeom:1993,Uhlmann:1995},  Hubner \cite{Hubner:1992,Hubner:1993a,Hubner:1993b}, 
Braunstein and Caves \cite{Braunstein_Caves:1994}
and Dittman \cite{Dittmann:1999},  and many others 
(see \cite{Jamiolkowski:2004,Zyczkowski_2ndEd:2020,Chien:2023} and references therein). The key concept for extending the above machinery from pure to mixed states, is to employ the purifications of $\rho$ at a point in $\M$ to pure states $\ket{\Psi}$ in $\F_\rho\subset\E$. The non-uniqueness of the purification in $\F_\rho$ lying above $\rho$ is given by the appropriate set of unitary transformations $\mathcal{U}$ of the fiber gauge group $\mathcal{G}$ such that $\ket{\Psi'}=\ket{\Psi}\,\mathcal{U}$ in $\F_\rho$ projects down to the same state $\rho = \pi(\ket{\Psi}) = \pi(\ket{\Psi'})$ in $\M$. Many of the concepts for the parallel transport (PT) of pure states have direct analogues for the case of mixed states, and the theoretical formulation well- (but not necessarily widely-) known. 
While many of the example calculations involving pure states are straightforwardly tractable \cite{Wilczek:1989,Bohm:2003,Jamiolkowski:2004,Zyczkowski_2ndEd:2020,Cafaro_Alsing:Grover:Geodesics:2020,Chien:2023,Alsing_Cafaro:GeometricAspects:2024,Alsing_Cafaro:BuresAndSjoqvistMetric:2023,Alsing_Cafaro:ComparingMetrics:2023,Cafaro_Alsing:Qubit:Geodesics:2023}, the case of mixed states is much more involved, with emphasis on the tractable case of qubits ($N=2^2$), and some results for qutrits ($N=3^2$), and even $n$-level systems (see references, page 236 of \cite{Jamiolkowski:2004}).
As recently reviewed by Hou \tit{et al.} \cite{Chien:2023}, for the case of pure states, the base space, projective Hilbert space $\M=\mathbb{C}P^{N-1}$, is a complex K\"{a}hler manifold with many nice properties. The `space' of all mixed states $\P$ is much more involved since involves flag manifolds (see Chapter 8 and 15 of Bengtsson and  Zyczkowski \cite{Zyczkowski_2ndEd:2020}  for further details, that will not be required here), whose gauge groups depend on the degree of degeneracy of the spectrum 
of the quantum state $\rho$ in $\M$. Denote as $\mathcal{D}^N_k$ the space of normalized density 
$N\times N$ density matrices with a fixed rank $k$ (number of non-zero eigenvalues), which is a manifold equipped with a Riemannian metric. Then $\P$ is not a manifold, rather a union of such manifolds given by  $\P=\bigcup_{k=1}^N\, \mathcal{D}^N_k$ which cannot be stitched together to construct a metric on $\P$ \cite{Chien:2023}. 
\bigskip

As an important example, the set of pure state density matrices is given by $\mathcal{D}^N_1$ of rank-1 projectors and is equivalent to $\M\simeq\mathbb{C}P^{N-1}$. In this case one takes as the bundle (or phase space) $\E$ the manifold of all normalized states $\mathcal{S}(\mathcal{H}) = S^{2N-1}$, the sphere of real dimension $2N-1$ (for the Hilbert space $\mathcal{H}=\mathbb{C}^N$), and obtains the base space manifold as the fibration (quotient space) $\M = S^{2N-1}/U(1)=\mathbb{C}P^{N-1}$ where the fibers are given by the  1-sphere $S^1$ of the $U(1)$ gauge group. That is, if $\ket{\psi(x)}\in\F_x$ is some pure state in the fiber 
$\F_x$ above a point $x\in \M$ in the base space (where $x=\{x_1, x_2,\ldots\}$ parameterizes $\M$), which projects down to the pure state $\rho(x)=\ket{\psi(x)}\bra{\psi(x)}$ at $x\in \M$, then $\ket{\psi(x)}\,e^{i\,\varphi(x)}\in\F_x$ also projects to $\rho(x)$.
\smallskip

The other important, tractable example primarily considered  in the literature 
 is the $N^2-1$ (real) dimensional space 
$\mathcal{D}^N_N$ of full-rank (and hence invertible) $N\times N$ matrices 
(note: $\mathcal{D}^N_k$ has real dimension $N^2-(N-k)^2-1$, with a proof given in 
Hou \tit{et al.} \cite{Chien:2023}).  $\mathcal{D}^N_N$ is not a complex K\"{a}hler manifold, and this characteristic embodies  the source of subtle differences between it and the pure state case $\mathcal{D}^N_1$ \cite{Chien:2023}. As is common in the literature, we will focus exclusively in this work on $\mathcal{D}^N_N$, the submanifold of all full-rank density matrices.
\smallskip

\begin{figure}[h]
\begin{center}
\includegraphics[width=4.25in,height=2.25in]{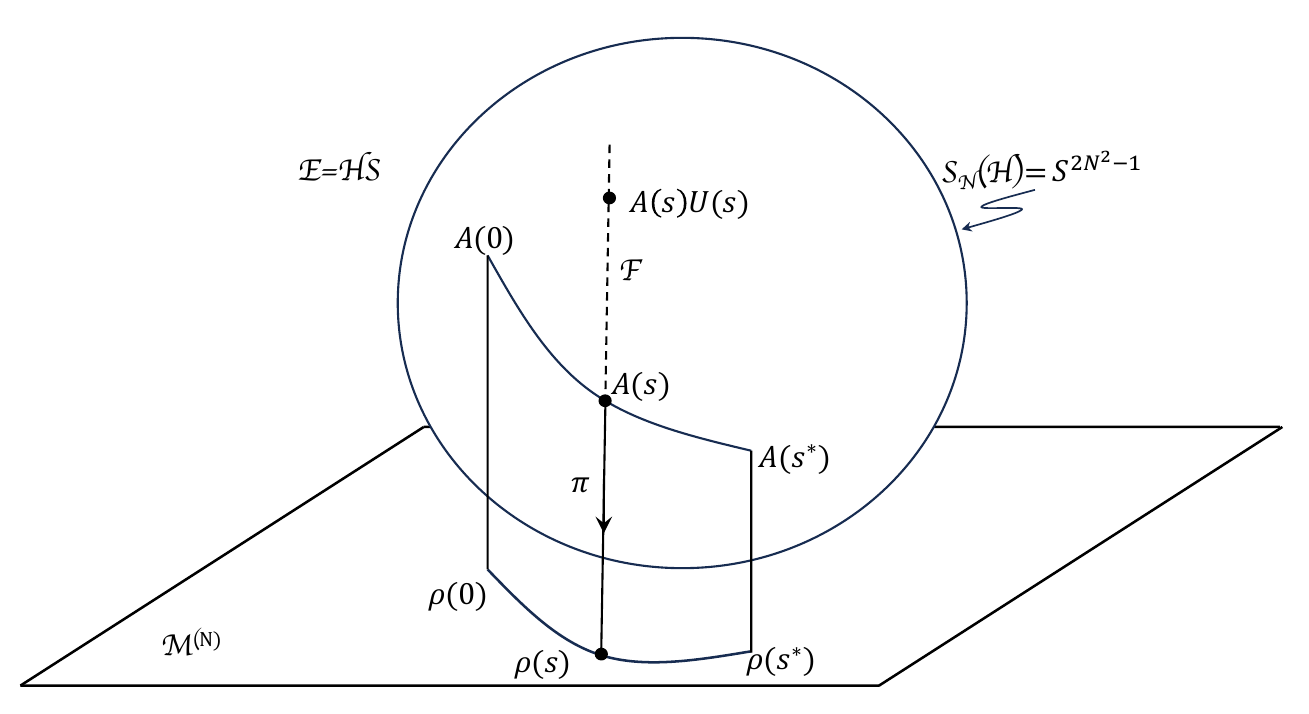}
\caption{Hilbert Schmidt Bundle $\E$ and relevant quantities.
Density matrix $\rho$ in base space $\MN$; purification $A$ in fiber $\F\subset\E$ above $\rho$.
Projection mapping $\{\pi: \F\subset\E\to\MN\, |\, \pi(A) = A\,A^\dag = \rho$\}.
$\mathcal{S}_{\mathcal{N}}(\mathcal{H})= S^{2N^2-1}\subset\E$ sphere of  normalized purified states.
}\label{fig:fiber:bunlde}
\end{center}
\end{figure}

It is well known from the works of Ulhmann that the arena for geometry of mixed states, denoted $\MN$, is given by the $N^2$ complex dimensional Hilbert-Schmidt space $\HS = \H\otimes\H^*$ (the algebra of complex matrices), 
where $\H$ is a complex $N$-dimensional Hilbert space (kets) and $\H^*$ is its dual (bras) such that any operator $A\in\HS$ can be written in the form $A= a\,\ket{P}\bra{Q} + b\,\ket{R}\bra{S}+\cdots$ provided enough terms are included in the sum \cite{Zyczkowski_2ndEd:2020}. $\HS$ is a Hilbert space itself when equipped with a Hermitian norm $\IP{A}{B}_{HS} = \Tr[A^\dag B]$. This scalar product gives rise to a Euclidean Hilbert-Schmidt distance $D^2_{HS}(A,B) = \thalf\,Tr[(A-B)^\dag\,(A-B)]$. 
Of relevance to the discussion of metrics and geodesics in $\MN$ will be the set $\P$ of positive operators $P\ge 0$, which can be written as $P=A\,A^\dag$, of unit norm with respect to the $\HS$ inner product. These operators can be considered as vectors in the unit sphere $\SN = S^{2 N^2-1}\subset \HS$, which plays a similar role to $\SH=S^{2N-1}$ in the above case of pure states, with the gauge group in the fiber now given by $U(N)$, replacing the phases $U(1)$ in the pure state case. Thus, the base space manifold of the Hilbert-Schmidt bundle $\E=\SN$ is given by the fibration $\MN = \SN/U(N) = \D^N_N$ \cite{Chien:2023}, in analogy to the case of pure states, when we restrict ourselves to the submanifold of unit trace positive operators $\Tr[\rho]=1$. Since $ \pi(A) = AA^\dag=\rho $ is invariant under gauge transformations $A\to A' = A\,U$ in the fiber (since $\rho'=A'\,A'^\dag = A\,U\,U^\dag A^\dag = \rho$) the pure state case of $\ket{\psi}\,e^{i\,\varphi}\in\E$ is `replaced' by $ A\,U\in\E$ in the case of mixed states.  
\Fig{fig:fiber:bunlde} is an illustrative depiction of the Hilbert Schmidt Fiber Bundle $\E$, and relevant objects and spaces discussed above.
%
\begin{table}[h]
\centering
\begin{tabular}
[c]{c|c|c|c|c}\hline\hline
\textbf{Fiber bundle quantity} & \textbf{Symbols}$_{\mathrm{pure}%
\text{\textrm{-}}\mathrm{states}}$ & \textbf{Elements}$_{\mathrm{pure}%
\text{\textrm{-}}\mathrm{states}}$ & \textbf{Symbols}$_{\mathrm{mixed}%
\text{\textrm{-}}\mathrm{states}}$ & \textbf{Elements}$_{\mathrm{mixed}%
\text{\textrm{-}}\mathrm{states}}$\\\hline
Total space, $\mathcal{E}$ & $S^{2N-1}$ & $\left\{  \left\vert \psi
\right\rangle \right\}  $ & $S^{2N^{2}-1}$ & $\left\{  A\right\}
$\\\hline
Base space, $\mathcal{M}$ & $S^{2N-1}/U(1)$ & $\left\{  \left\vert
\psi\right\rangle \left\langle \psi\right\vert =\rho\right\}  $ &
$S^{2N^{2}-1}/U(N)$ & $\left\{  AA^{\dagger}=\rho\right\}  $\\\hline
Fiber space, $\mathcal{F}_{\mathcal{E}}$ & $\mathcal{F}_{S^{2N-1}}$ &
$\left\{  \left\vert \psi\right\rangle e^{i\varphi}\right\}  $ &
$\mathcal{F}_{S^{2N^{2}-1}}$ & $\left\{  AU\right\}  $\\\hline
Structure group, $\mathcal{G}$ & $U(1)$ & $\left\{  e^{i\varphi}\right\}  $ &
$U(N)$ & $\left\{  U\right\}  $\\\hline
Projection, $\pi$ & $\pi:S^{2N-1}\rightarrow S^{2N-1}/U(1)$ & $\pi\left(
\left\vert \psi\right\rangle \right)  =\left\vert \psi\right\rangle
\left\langle \psi\right\vert =\rho$ & $\pi:S^{2N^{2}-1}\rightarrow
S^{2N^{2}-1}/U(N)$ & $\pi\left(  A\right)  =AA^{\dagger}=\rho$\\\hline
\end{tabular}
\caption{Schematic summary of the most important fiber bundle quantities for
pure and mixed quantum states scenarios.}%
\label{table:List:of:Symbols}
\end{table}
In Table~\ref{table:List:of:Symbols} we present a list of symbols of the various geometric quantities used throughout this work.
\smallskip

A note on the representation of $A$. So far we have discussed representing pure states in $\E$ by positive operators $A$. A particular representation is given by $A = \sqrt{\rho}$, which, since it is valid everywhere in $\E$ is called a \tit{global section} (a mapping $\pi^{-1}(\rho) = \sqrt{\rho}$ from $\MN\to\F\subset\E$  for every point along a curve in the base manifold ).  If one employs a spectral representation of $\rho$ and 
chooses $A'=\sqrt{\rho}=\sum_i\,\sqrt{\lambda_i}\,\ket{i}\bra{i}$ for a given fiber, then any other element in that fiber can be represented by $A=\sum_i\,\sqrt{\lambda_i}\,\ket{i}\bra{i}\,U$ for $U\in U(N)$, which one might call a `state-matrix." One can define an associate `state-vector' 
$\ket{A} = \sum_i\,\sqrt{\lambda_i}\,\ket{i}\otimes U^T\ket{i}$  by taking the ordinary transpose $T$ 
of `$\ket{i}\,U$'  in $A$ \cite{Chien:2023}. 
\smallskip

The above is consistent with another common representation of the purification of $\rho$ by a state vector (see e.g. Wilde \cite{Wilde:2017}) as 
$\ket{\psi_\rho} = (U_E\otimes\sqrt{\rho_S})\,\ket{\G}_{ES}$, where $\ket{\G}_{ES}=\sum_i \ket{i}\otimes\ket{i}$ is the unnormalized maximally entangled  bipartite (Environment-System) state-vector. By the (straightforwardly proven) `kickback theorem' \cite{Wilde:2017} the bipartite state 
$\ket{\G}$ has property $(U\otimes V)\ket{\G} = (I\otimes V\,U^T\ket{\G}=(U\,V^T)\otimes I)\ket{\G}$ , where without loss of generality, one can always assume the dimension of the environment ($K \ge N$) to be equal to the dimension of the system. Thus, one can then write always 
$\ket{\psi_\rho} =  (U\otimes\sqrt{\rho})\,\ket{\G} 
                         = (U\otimes I)\,\sum_i\,\ket{i}\otimes \sqrt{\lambda_i}\,\ket{i} 
                         = \sum_i\,\sqrt{\lambda_i}\,\ket{i}\otimes U^T\ket{i} = \ket{A}
$.
The effective projection map $\pi$ is the partial trace over the environment:  
$\pi({\ket{\psi_\rho}}) = \Tr_E[\ket{\psi_\rho}\bra{\psi_\rho}] = \rho$. The state-vector representation (vs the state-matrix form) is the more familiar association with the phrase 
\tit{a pure state purification of} $\rho$, which is the Schmidt decomposition of a bipartite state in a higher dimensional (total space) $\E$, whose reduced density matrix is~$\rho$.
\smallskip

Lastly, we could also think of the purification $A$ in the following way \cite{Zyczkowski_2ndEd:2020} (relaxing the requirement that $A$ be represented by a square matrix). Let the pure state $\ket{\psi_\rho}_{SE}\in\E$, the purification of the density matrix $\rho\in\M$, be given by $\ket{\psi_\rho}_{SE} = \sum_{n=0}^{N}\sum_{k=0}^{K\ge N} A_{nk}\,\ket{n}_S\otimes\ket{k}_E$, where $S$ stands for \tit{system} living in the base space $\M$, and $E$ stands for the environment, and for convenience we take $K\ge N$ without loss of generality, so that $A$ is an $N\times K$ matrix. The system density matrix  is given by
$\rho_S = \Tr_E\big[\ket{\psi_\rho}_{SE}\bra{\psi_\rho}\big] 
= \sum_{n,n'}^{N} \left(\sum_k A_{n k}\,A^*_{n' k}\right)\,\ket{n}_S\bra{n'}\equiv 
\sum_{n,n'}^{N} (A\,A^\dag)_{n n'}\,\ket{n}_S\bra{n'}$. Thus, $\pi(A) = \rho_S \equiv  A\,A^\dag$ is just the representation of $\rho_S$ by the positive matrix $A\,A^\dag$ derivable from the $N\times K$ matrix $A$ 
describing the purification state $\ket{\psi_\rho}_{SE}$.
In the following, a specific representation is not required in order to derive general results, but instead will be invoked when specific examples are discussed.
\bigskip

The outline of this paper is as follows:
\smallskip

\indent In \Sec{sec:PTofHSB} we briefly review the condition for parallel transport in the Hilbert-Schmidt bundle $\E$, which gives rise to the horizontal lift condition (HLC). The HLC  defines horizontal subspace of tangent vectors in $\E=\SN$ orthogonal to the vertical subspace of tangent vectors that lie along the fibers $\F$ above the base space 
$\MN$. Since geodesics in $\SH=S^{2N^2-1}$ are simply great circles, (which project down to geodesics in $\MN$) one can develop a simple formula for $A(s)$, where $s$ is an affine parameter along the geodesic curve.
\smallskip

In \Sec{sec:Bures} we review the Bures distance and Bures angle between two mixed states in $\E$, and define and outline the proof of the Uhlmann fidelity, which measures the transition probability between the states.
\smallskip


In \Sec{sec:Bures:metric}  we review the Bures metric (infinitesimal Bures distance) and explore the HLC which leads to the (in general) non-unitary evolution equation for $\rho$ in terms of the symmetric logarithmic derivative, i.e. $\dot{\rho} = G\,\rho + \rho\,G$ for $G$ Hermitian. We explore solutions for $G$ in terms of $\rho$ for arbitrary dimension $N$, for both non-unitary and unitary geodesics  
between states.
\smallskip

In \Sec{sec:HL:geodesic} we exploit the property of the geometric mean operator 
$M = \rho_1^{-1/2}\, \sqrt{\rho_1^{1/2}\rho_2\rho_1^{1/2}}\,\rho_1^{-1/2}$, namely $A_2 = M\,A_1$ where
$\rho_1=A_1\,A_1^\dag$ and $\rho_2=A_2\,A_2^\dag$, to develop a tractable formula for the intermediate states $\rho(s)$ along the geodesics such that $\rho_1=\rho(0)$, $\rho_2=\rho(s^*)$, and 
$\cos(s^*) = \sqrt{F(\rho_1,\rho_2)} = \Tr[M\,\rho_1]$ is the Uhlmann root fidelity between the initial and final density matrices. 
\smallskip

In \Sec{sec:Examples} we present three examples illustrating the formulas developed in
\Sec{sec:HL:geodesic}.
 The first example is the geodesic between the maximally mixed state and a pure state, in arbitrary dimensions. The second example is the geodesic
between Werner states
$\rho(p) = (1-p)\,\Id/N + p\,\ket{\Psi}\bra{\Psi}$
with 
$\ket{\Psi} = \{\ket{GHZ}, \ket{W}\}$
in dimension $N=2^3$. For the latter, we compare the results in the limit $p\to1$ to the infinite number of possible geodesics between the orthogonal pure states $\ket{GHZ}$ and $\ket{W}$.
Lastly, we analytically computed the geodesic orbit between two given qubit mixed states 
within the Bloch sphere for dimension $N=2$.
\smallskip

In \Sec{sec:Summary:Conclusion} we briefly summarize our main results 
and present our conclusions. Lastly, we indicate directions for future research.

\section{Parallel Transport and the horizontal lift condition in the Hilbert-Schmidt bundle $\HS=\E$}\label{sec:PTofHSB}
In this section we review the definitions and derivation of the horizontal lift $A\in \E$ of the curve 
$\rho = \pi(A) = A A^\dag\in\MN$ which defines the condition for parallel transport for a curve in $\E$, i.e. one that winds its way through $\E$ perpendicular to the fibers $\F\subset\E$.

\subsection{Defining the \tit{horizontal subspace $H(\E)$} }\label{GQS:Sec:9.3}
We want to consider the $\HS$ bundle as a \tit{real} vector space of dimension $2 N^2$, so we adopt a metric (\tit{inner product}) of the form \cite{Zyczkowski_2ndEd:2020}
\bea{GQS:9.21}
X\cdot Y &=& \dhalf \left( \braket{X,Y} + \braket{Y,X} \right)  = \dhalf\,\tr{X^\dag Y + Y^\dag X}, \label{GQS:9.21:line1} \\
&=&  \dhalf\,\tr{X^\dag Y + (X^\dag Y)^\dag},\no
&=& \re{\tr{X^\dag Y}}
\eea
(Because we are in a vector space, the tangent spaces can be identified with the space itself).
A matrix (vector in $\HS$) in the Hilbert-Schmidt bundle will only project down to a properly normalized density matrix $\rho\in \MN$ \tit{if and only if} it sits on the unit sphere $S^{2 N^2-1}$ in $\HS=\E$.

As discussed in \Sec{sec:intro}, a \tit{connection} is equivalent to a decomposition of the bundle tangent space into a vertical and horizontal component $\TE=\VE\oplus\HE$. The \tit{vertical tangent vectors} are easy to define, since by definition they point along the fibers $\F\subset\E$.
The unitary $U(N)$ acts to move a purification $A$ around within a fiber, so that a curve, parameterized by $t$ in the fiber is given by
\be{GQS:9.22}
A U(t) = A e^{i H t}, \quad H^\dag = H,
\ee
where we have written $U$ as the exponential of some Hermitian matrix $H$ (which we can always do).
The \tit{tangent vector} $\dA$ to this curve in the fiber is defined as
\be{Adot}
\dA = \dfrac{d}{dt} \Big( A e^{i H t} \Big)_{t=0} = i A H \in\VE, \quad\;  \trm{vertical tangent vector}.
\ee
The \tit{horizontal vectors} $\HE$ must be defined somehow, and this is accomplished by \tit{defining}  $\HE$ to be those vectors which are \tit{orthogonal to} $\VE$ \tit{with respect to the metric} \Eq{GQS:9.21} in $\HS=\E$.
For a arbitrary horizontal vector $X\in\HE$ this means, by \Eq{GQS:9.21}, that we must have
\bsub
\bea{GQS:9.23}
0 = X\cdot(i A H) &=& \dhalf \left( \tr{X (i A H)} + \tr{\big( X (i A H)\big)^\dag} \right), \label{GQS:9.23:line1}\\
&=&  i\,\dhalf \tr{\left( X^\dag A - A^\dag X\right) H}. \label{GQS:9.23:line2}
\eea
\esub
Since this must be true for \tit{any} horizontal vector $X\in\HE$ and arbitrary path in $\F$, i.e. for \tit{arbitrary} $H$, \Eq{GQS:9.23:line2} implies that \tit{a horizontal vector at the point $A$} is given by the \tit{horizontal lift condition} (HLC)
\bsub
\bea{GQS:9.24}
X\in\HE &\Leftrightarrow& X^\dag A - A^\dag X = X^\dag A - \left(X^\dag A\right)^\dag = 0, \label{GQS:9.24:a} \\
&\Leftrightarrow& \left(X^\dag A\right)^\dag = X^\dag A \quad \Rightarrow \quad X^\dag A\; \trm{is Hermitian}. \label{GQS:9.24:b}
\eea
\esub
Armed with this definition of what it means to be a horizontal vector, we can now define the \tit{horizontal lift} of a curve in the base manifold $\MN$ to the bundle $\E$.

Let $A(s)$ be a \tit{curve} in the bundle $\E$ (parameterized by an affine parameter $s$) such that $A(s) A^\dag(s) = \rho(s)$, with \tit{tangent vector} $\dA(s)$. We say this tangent vector is \tit{horizontal} if it satisfies the HLC above. 
Setting $X\to\dA$ in \Eq{GQS:9.23} we have
\be{GQS:9.25}
\dA^\dag A = A^\dag \dA  \equiv \left( \dA^\dag A\right)^\dag \quad \Rightarrow \quad \dA^\dag A\; \trm{is Hermitian}.
\ee
Note that $A^\dag\,\dot{A}$ is analogous to the amplitude $\IP{\Psi}{\dot{\Psi}}$ so that \Eq{GQS:9.25} is the mixed state analogue of the HLC for the $U(1)$ bundle for pure states, namely that the connection
$\mathcal{A} = \trm{Im}[\IP{\Psi}{\dot{\Psi}}] = 0$.
\bigskip

It is trivial to see that \Eq{GQS:9.25} can be satisfied by
\be{GQS:9.26}
\dA = G A, \quad G^\dag = G,
\ee
for $G$ Hermitian, since both sides of  \Eq{GQS:9.25} become $A^\dag G A$, using $\dA^\dag = (G A)^\dag = A^\dag G$.
Now a quick calculation reveals that 
\bea{GQS:9.27}
A(s) A^\dag(s) = \rho(s) \quad \Rightarrow \quad \dot{\rho}(s) &=& dA(s) A^\dag(s) + A(s) \dA^\dag(s)  \no
&=& G(s) A(s) A^\dag(s) + A(s) A^\dag(s) G, \no
&\equiv& G(s) \rho(s) + \rho(s) G(s).
\eea
We recognize the last term as the \tit{symmetric logarithmic derivative} 
that plays a prominent role in the discussion of  the Quantum Fisher Information (QFI)  \cite{Braunstein_Caves:1994,BAG_Parity:2021}.
\Eq{GQS:9.26} has the \tit{formal} solution (\cite{Louisell:1973}, page 65)
\bsub
\bea{A:formal:soln}
A(s) &=& \T\left[ e^{\int_0^s G(s') ds'}\right] A(0),   \qquad\quad
\T[G(s) G(s')] = 
\begin{cases}
G(s) G(s'), \; s > s' \\
G(s') G(s), \; s' > s
\end{cases} , \label{A:formal:soln:line1} \\
&=& 
\T\,\left[
\Id + \sum_{n=1}^\infty \dfrac{1}{n!}\, \int_{0}^{s} ds_1  \int_{0}^{s} ds_2 \ldots  \int_{0}^{s} ds_n \, G(s_1) G(s_2)\cdots G(s_n)
\right]\,A(0), \label{A:formal:soln:line2} \\
&\equiv& 
\left[
\Id + \sum_{n=1}^\infty  \int_{0}^{s} ds_1  \int_{0}^{s_1} ds_2 \ldots  \int_{0}^{s_{n-1}} ds_n \, G(s_1) G(s_2)\cdots G(s_n)
\right]\,A(0),
\label{A:formal:soln:line3}
\eea
\esub
where $\T$ is the \tit{time ordering operator} given by the expression on the right in \Eq{A:formal:soln:line1}, and explicitly exhibited in the 
nested, time-ordered integrals in \Eq{A:formal:soln:line3}.
Note that \Eq{A:formal:soln:line1}, \Eq{A:formal:soln:line2} and  \Eq{A:formal:soln:line3} are most often encountered in QM in the context of \tit{perturbation theory} where $G\to -i\,H$ for Hamiltonian $H^\dag =H$. However, recall $G^\dag = G$, but $( -i\,H)^\dag = i\,H$ is \tit{anti-Hermitian}. 
Thus, in general  \Eq{A:formal:soln:line1} is \tit{not} a unitary evolution. 
In general, it is difficult to find explicit expressions for $G$ except in special case, e.g. $N=2$, or in terms of the eigenvalues of $\rho$ (see \cite{BAG_Parity:2021} and references therein).
Dittmann \cite{Dittmann:1999}  has given explicit, though complicated series expressions for $G$ that can be obtained \tit{without} the need to diagonalize $\rho$, i.e. in terms of invariants that arise out of the characteristic polynomial for $\rho$, 
i.e. $0 = \trm{det}(\lambda\Id - \rho) = \l^N -\S_1 \l^{N-1} + \S_2 \l^{N-2} - \cdots + (-1)^N \S_N $ where 
$\S_1 = \sum_i \l_i = \tr{\rho}$, $\S_2 = \sum_{i<j} \l_i \l_j = \thalf(\S_1\tr{\rho} - \tr{\rho^2})$, $\S_3 = \sum_{i<j<k} \l_i \l_j \l_k$, \ldots are symmetric invariant polynomials of the eigenvalues of $\rho$, and iteratively $\S_k = \tfrac{1}{k} \sum_{j=1}^{k} (-1)^{j-1}\S_{k-j} \tr{\rho^k}$
(see Section 8.1 of \cite{Zyczkowski_2ndEd:2020},  and page 297 Eguchi \tit{et al.} \cite{Eguchi:1980} ).
\bigskip

The takeaway point here is that 
\begin{itemize}
\item The horizontal lift $A(s)\in\E$ of a curve in the base manifold $\rho(s)\in\M$ is \tit{uniquely determined} by the horizontal condition \Eq{GQS:9.25} and \Eq{GQS:9.26}, which is equivalent to the specification of a \tit{connection} on $\HS=\E$.
\item The HLC \Eq{GQS:9.24:a} and \Eq{GQS:9.24:b} \tit{is} the \tit{parallel transport} (PT) rule in $\E$.
\end{itemize}
\bigskip
In \Sec{sec:Bures:metric}  we will examine expressions for $G$ in terms of $\rho$, revisiting \Eq{GQS:9.27} for general $N$. Ulhmann \cite{Uhlmann:DenOprs:DiffGeom:1993} has shown that unitary evolution can be expressed by 
$G=i\,[\rho, Y]$ for $Y=Y^\dag$ Hermitian, which reproduces Louiville's equation 
$i\,\dot{\rho}=[H,\rho]$ with $H=\rho\,Y + \rho, Y$, which is easily proven by substituting the expression for $G$ into \Eq{GQS:9.27} and rearranging terms.

\section{The Bures distance, angle and Uhlmann Fidelity}\label{sec:Bures}
Before we examine expressions for G, we first review the Bures distance, angle and Uhlmann fidelity between two quantum states.

\subsection{The Bures distance $D_B(\rho_1,\rho_2)$, Bures angle $D_A(\rho_1,\rho_2)$ and the root Fidelity  $\rootF$ }
In addition to the connection (which defines the HLC) in the previous section, the $\HS$ bundle also admits a natural metric on the space of density matrices $\rho\in\MN$ ($\tr{\rho}=1$) known as the \tit{Bures metric}, which lives on the positive cone $\P=\MN$ of operators $P\ge 0$ ($\tr{P}\ne1)$ on $\H$, which is our base manifold $\MN$.
\smallskip

In this section we will treat $\rho$ as \tit{any} positive operator (i.e. $\rho\in\P$) and allow $\tr{\rho}\ne 1$.
The \tit{purification of $\rho$} is given by $A\in\HS=\E$ such that $\rho = A A^\dag$, where $A$ is regarded as a \tit{vector} (pure state) in $\HS$.

In the bundle space $\HS=\E$ we have a natural notion of distance given by the Euclidean (Hilbert-Schmidt) distance defined by 
\be{GQS:9.28}
d_B^2(A_1,A_2)  = ||A_1-A_2||^2_{HS} = \tr{A_1 A_1^\dag + A_2 A_2^\dag  - (A_1 A_2^\dag + A_2 A_1^\dag )}.
\ee
If \tit{additionally} $A_1$ and $A_2$ lie in the unit sphere $\SN(\H)=S^{2 N^2-1}$ (i.e. we additionally impose that $\tr{\rho_1} = \tr{\rho_2}=1$) then we have \tit{another natural distance}, namely the \tit{geodesic distance} $d_A$ (on the sphere) given by
\be{GQS:9.29}
\cos d_A = \dhalf \tr{A_1 A_2^\dag + A_2 A_1^\dag },
\ee
namely, one half the terms with a ``-" sign in \Eq{GQS:9.28}.
Note that while $d_B^2(A_1,A_2)$ measures the straight chord Euclidean distance, 
the  distance $D_B(\rho_1, \rho_2)$  measures the length of a curve  in $\MN$
 that projects down from $\E$, in its entirety, to a curve traced out by a density matrix $\rho\in \MN$.
As discussed earlier, the natural definition of a distance between two density matrices $\rho_1$ and  $\rho_2$ is the length of the shortest path that connects the two fibers lying over these density matrices
\begin{itemize}
\item The distance between two density matrices $\rho_1$ and  $\rho_2$ in base manifold $\M$ is \tit{defined} as the \tit{shortest distance} along a path in the bundle between their purifications $A_1$ and $A_2$ that lie in the fibers above them. This shortest path lies along the geodesic defined by the HLC.
\end{itemize}
Thus, either working with $d_B$ or $d_A$, the goal is to calculate the Uhlmann \tit{root fidelity} $\rootF$ defined by
\bsub
\bea{GQS:9.30}
\rootF &=& \dhalf \trm{max} \tr{A_1 A_2^\dag + A_2 A_1^\dag}, \\
&=& \trm{max} \re{\tr{A_1 A_2^\dag}}, \no
&=&  \trm{max} \left| \tr{A_1 A_2^\dag} \right|,
\eea
\esub
where the optimization is taken with respect to \tit{all possible purifications} of $\rho_1$ and  $\rho_2$.
This in turn defines the Bures  distance $D_B$ and the Bures angle $D_A$
\bsub
\bea{GQS:9:31:9:32:again}
D_B^2(\rho_1, \rho_2) &=& \tr{\rho_1} + \tr{\rho_2} - 2\,\rootF,
\label{GQS:9.31:again} \\
\cos\,D_A(\rho_1, \rho_2) &\equiv& \rootF  = \tr{\roottauopr\,}.                                   \label{GQS:9.32:again}
\eea
\esub
Again, the Bures angle $D_A$ measures the length of a curve in in the bundle $\MN\subset\HS=\E$, while the Bures distance measures the length of a curve within the positive cone (base manifold) $\P=\MN$. They are both Riemannian metrics, and more importantly, they are both 
\tit{monotonically decreasing} functions of the root fidelity. This means under any quantum  operation  
$\N$, such that $\rho\to\rho' = \sum_i E_i \rho E_i^\dag$ with $\sum E^\dag_i E_i = \Id$  we have
$F(\N(\rho_1), \N(\rho_2)) \ge F(\rho_1,\rho_2)$  \cite{NC:2000,Wilde:2017}. This last statement says that for \tit{any} physical process (unitary (trace preserving) or otherwise (non trace preserving, e.g. measurements)) the \tit{fidelity never decreases}, i.e. it can only stay the same or increase, meaning that the states are becoming \tit{more identical} or equivalently \tit{less distinguishable}, under the process. By \Eq{GQS:9.31:again}, this means that the Bures distance can only \tit{decrease}, (since the root fidelity enters $D^2_B$ with a \tit{minus} sign. Again, smaller Bures distance implies the states are becoming less distinguishable).
\bigskip

\subsection{Proof of Uhlmann's fidelity theorem}
The proof of Uhlmann's fidelity theorem is well known, and clear expositions are given in many places, in particularly Wilde \cite{Wilde:2017}, and Bengtsson and Zyczkowski \cite{Zyczkowski_2ndEd:2020}.
The proof is amazingly simple and illuminating, and so for completeness, 
we outline its essential elements below.
The theorem is stated as follows

\flushleft{\bf Uhlmann's fidelity theorem:} \tit{The root fidelity, defined as the maximum of $\tr{A_1 A_2^\dag}$ over all possible purifications of two density matrices $\rho_1$ and  $\rho_2$, is}
\be{GQS:9.33}
\rootF = \tr{\, \big|\rootrhoone \rootrhotwo\,\big|\,} = \tr{\roottauopr}.
\ee
To prove this non-intuitive looking formula let us use the polar decomposition \cite{NC:2000} to write the most general purification in the fibers above the density matrices as
\be{GQS:9.34}
A_1 = \rootrhoone\, U_1, \quad A_2 = \rootrhotwo\, U_2,
\ee
where $U_1$ and $U_2$ are some unitary operators that move states around within the fibers above $\rho_1$ and  $\rho_2$, respectively.
Then, we have
\be{GQS:9.35}
\tr{A_1 A_2^\dag} = \tr{ \rootrhoone U_1 \,U_2^\dag \rootrhotwo} = \tr{\rootrhotwo  \, \rootrhoone \,  U_1 U_2^\dag}.
\ee
Now the  operator $\rootrhoone  \rootrhotwo$ is positive, but not Hermitian. Note that \tit{if}  $\rho_1$ and  $\rho_2$ were diagonal matrices, then modulo the unitaries (which we could take to be unity) \Eq{GQS:9.34} would yield $\tr{\rootrhoone  \rootrhotwo} \xrightarrow{\trm{diagonal}} \sum_i p_i q_i$ $\equiv F_{classical}$ where the last expression is the \tit{classical fidelity} or  \tit{classical Bhattacharya distance} between two probability distributions $\{p_i \}$ and $\{q_i \}$ associated with the diagonal entries of $\rho_1$ and  $\rho_2$. However, this is a very special situation.
In general, we proceed by taking one last polar decomposition, that of  
\bsub
\bea{GQS:9.36}
\rootrhoone  \rootrhotwo &=& |\rootrhoone  \rootrhotwo| \, V, \quad V^\dag = V, \\
 |\rootrhoone  \rootrhotwo| &\equiv& \sqrt{(\rootrhoone  \rootrhotwo) (\rootrhoone  \rootrhotwo)^\dag} = \roottauopr.
\eea
\esub
Now define the unitary $U\equiv V\,U_2\,U_1^\dag$, so that the final task is to maximize the quantity
\be{GQS:9.37}
\underset{U}{\trm{max}}\,\tr{\, |\rootrhoone\,\rootrhotwo| \, U \, }, 
\ee
where the max is taken over all unitary operators $U$.
This maximum occurs when $U=\Id$,
which follows from the definition of the $L_1$ norm via the definition 
$\trm{max}_{U} \,\tr{\, M\, U \,} \equiv||M||_1\equiv$$\tr{\, |M| \, }$, 
with $|M| = \sqrt{M\,M^\dag}$, 
valid for any matrix $M$ \cite{Wilde:2017}.
%
%
Hence, the proof of Uhlmann's fidelity theorem is complete, and subsequently the definitions of the Bures distance and Bures angle.
It is curious that for full rank (hence, invertible) density matrices, the naive guess for
$U_1\,U_2^\dag = \sqrt{\rho_1^{1/2}\,\rho_2\,\rho_1^{1/2}\,}\,\rho_1^{-1/2}\,\rho_2^{-1/2}$ in
\Eq{GQS:9.35}, which through simple multiplication yields the expression for the fidelity
$\rootF = \Tr\left[\sqrt{\rho_1^{1/2}\,\rho_2\,\rho_1^{1/2}}\right]$, is actually unitary (see Appendix \ref{app:A} for details).

\section{The Bures metric}\label{sec:Bures:metric}
With the above results in hand, we can now construct the Riemannian metric defined by the Bures distance.

\subsection{The Riemannian metric defined by the Bures distance}
Instead of $\dA$ let us write equivalently $dA$, which projects down to $d\rho$. The length squared of $d\rho$ is \tit{defined} by
\be{GQS:9.41}
ds^2 = \trm{min} \tr{dA \,dA^\dag},
\ee
where the minimum is taken over \tit{all} vectors $dA$ that project down to $d\rho$. From the previous discussion, we have seen that this minimal distance is achieved \tit{when $dA$ is a horizontal vector}, which occurs \tit{if and only if} $dA = G A$, where $G^\dag = G$ is a Hermitian matrix. For $\rho>0$ (strictly positive) this $G$ is defined uniquely by 
$d\rho = d(A A^\dag)$ $= dA \, A^\dag + A \, dA^\dag$ $= (G A) A^\dag + A (A^\dag G)$ $= G \rho + \rho G$.
\be{GQS:9.42}
dA = G\,A \Rightarrow d\rho = G\,\rho + \rho\,G, \quad G^\dag = G.
\ee
Substituting $dA = G A$ into \Eq{GQS:9.41} yields
\be{GQS:9.43}
ds^2 = \tr{(G A)\,(A^\dag G)} = \tr{G \rho G} = \dhalf\,\tr{G\,d\rho},
\ee
where in the last equality we have noted that $\tr{G d\rho} = \tr{G (G\rho + \rho G)} = 2\,\tr{G \rho G}$.
Note, that the expression on the right hand side of \Eq{GQS:9.43} is also called the \tit{Quantum Fisher Information} (QFI) $\F_{QFI}$ (see \cite{BAG_Parity:2021}), and
so we see that 
\bsub
\bea{ds:QFI:Fidelity}
ds^2_B &=& \F_{QFI} = \tr{G \rho G} = \dhalf\,\tr{G\,d\rho}   \label{ds:QFI:Fidelity:line1} \\
 \Rightarrow 
 s_B(\rho_1,\rho_2) &=& \int_0^\s \sqrt{ \F_{QFI}(\s')} \, d\s' = \cos D_A(\rho_1,\rho_2) = \rootF  = \tr{\roottauopr},  \label{ds:QFI:Fidelity:line2}
\eea
\esub
with $\rho_1=\rho(0)$ and $\rho_2=\rho(\s)$, and 
where the second line follows since $s_B(\rho_1,\rho_2)$ is just the total arclength between $\rho_1$ and $\rho_2$, which from before, is the Bures angle given by  $\cos D_A(\rho_1,\rho_2) = \rootF$. The key takeaway point here is
\begin{itemize}
\item The root QFI  $\sqrt{\F_{QFI}}(\rho_1,\rho_2)$ is the infinitesimal squared distance measure (i.e. the Bures metric $ds^2_B(\rho_1,\rho_2)$) of the root Quantum Fidelity $\rootF$ between two close quantum states $\rho_1$ and $\rho_2=\rho_1+d\rho$.
\end{itemize}
\bigskip

The derivation of the Bures metric follows from expanding the root fidelity $\rootF$ in  the Bures distance $D^2_B(\rho_1,\rho_2)$ in \Eq{GQS:9.31:again} with $\Tr[\rho]=1$, using $\rho_1\to\rho$ and 
$\rho_2\to \rho+t\,d\rho$ 
(i.e. $\sqrt{F}\to \sqrt{F}(t)\defn \sqrt{\rho^{1/2}\,(\rho + t\,d\rho) \, \rho^{1/2}}$),
and expanding to second order in $t$, and evaluating at $t=0$.
The derivation is due to H\"{u}bner  \cite{Hubner:1992}, (further details can also be found in \cite{Zyczkowski_2ndEd:2020,Alsing_Cafaro:BuresAndSjoqvistMetric:2023,Chien:2023}), 
and yields
\be{Bures:metric}
ds^2_{Bures}(\rho, \rho+d\rho) =  \half\,\sum_{i,j}\,\frac{|\EV{i}{d\rho}{j}|^2}{\lambda_i+\lambda_j}
=\half\,\sum_{i,j}\,\frac{(\sqrt{\lambda_i}+\sqrt{\lambda_j})^2}{\lambda_i+\lambda_j}\,
|\EV{i}{d\sqrt{\rho}}{j}|^2,
\ee
where one has employed a spectral representation of $\rho=\sum_{i=1}^N\, \lambda_i \, \ket{i}\bra{i}$. In the second equality above one has used $d\rho = (d\sqrt{\rho})\,\sqrt{\rho} +\sqrt{\rho}\, (d\sqrt{\rho})$ so that 
$\EV{i}{d\rho}{j} = (\sqrt{\lambda_i}+\sqrt{\lambda_j})\,\EV{i}{d\sqrt{\rho}}{j}$.
For the case of a mixed state qubit $N=2$, with 
$\rho = \thalf( \Id + \bfx\cdot\bfsigma)$ one obtains  \cite{Jamiolkowski:2004}
$ds^2_B = \tfrac{1}{4}\left( (d\bfx)^2  + \dfrac{(\bfx\cdot d\bfx)^2}{1-|\bfx|^2}\right)$, which upon defining $x^2_4 \defn 4\det(\rho) = 1-|\bfx|^2$ becomes 
$ds^2_B = \tfrac{1}{4}\sum_{i=1}^{4} (dx_i)^2$, the metric on the 3-sphere $S^3$. Thus, the mixed state Bloch ball can be isometrically embedded into the hemisphere of $S^3$ of radius $\thalf$, defined by $x_4\ge 0$. The pure states ($|\bfx|=1$) are defined on the boundary, the equator $S^1$, and the maximally mixed state $\Id/2$ ($|\bfx|=0$) is at the `north pole.' 

\subsection{Linear equations for $G$ in terms of $\rho$ for arbitrary $N$: general, non-unitary evolution}
The parallel transport (HLC) $\dot{A} = G A\in\E$, $G^\dag = G$ 
leads to the (in general, non-unitary) evolution equation 
$\dot{\rho} = G\,\rho + \rho\,G, \quad G^\dag = G$ for $\rho\in\MN$
in \Eq{GQS:9.27} and \Eq{GQS:9.42}.
In this section we derive linear equations for $G$ in terms of $\rho$.
Since $\rho$ depends on the geodesic path $\bfx$, such expressions for $G$ imply that 
one must first predetermine the path, say by solving the (Euler-Lagrange) geodesic equations from the metric $ds^2_B$ (e.g. see \cite{Alsing_Cafaro:ComparingMetrics:2023}).
\bigskip

Using the defining representation of the Lie algebra of $SU(N)$ in terms of $N\times N$ traceless Hermitian matrices $\{\sigma_i\}|_{i=1,\ldots,N^2-1}$ as (see Appendix B of \cite{Zyczkowski_2ndEd:2020})
we have the properties
\bsub
\bea{Lie:Algebra:SUN}
\sigma_i\,\sigma_j &=& \frac{2}{N}\,\delta_{ij} + d_{ijk}\,\sigma_k + i\,f_{ijk}\,\sigma_k, \label{Lie:Algebra:SUN:a} \\
\left[\sigma_i, \sigma_j\right] &=& 2\,i\,f_{ijk}\,\sigma_k,  \quad
\{\sigma_i\,\sigma_j\} = 4\,\delta_{ij}\, \Id/N + 2\,d_{ijk}\,\sigma_k, \label{Lie:Algebra:SUN:b} \\
\Tr[\sigma_i\,\sigma_j] &=& 2\, \delta_{ij}, \hspace{0.45in} 
(\sigma_i)_A^{\; B}\,(\sigma_i)_C^{\; D} = 2\, \delta_A^{\; D}\,  \delta_B^{\; C}
- \frac{2}{N}\,\delta_A^{\; B}\,  \delta_C^{\; D}, \label{Lie:Algebra:SUN:c}
\eea 
\esub
where $A,B,C,D=1,\ldots,N$, 
$f_{ijk}$ are the totally antisymmetric structure constants characterizing the Lie algebra, 
the $d_{ijk}$ are totally symmetric symbols, which are representation dependent, and the rightmost expression in \Eq{Lie:Algebra:SUN:c} is a statement of completeness. For $N=2$, $d_{ijk}=0$, $f_{ijk}=\epsilon_{ijk}$, the above reduce to the standard $2\times 2$ Pauli matrices.
Let us now define an arbitrary Hermitian matrix with $g_0=\Tr[G]$ in terms of the basis
$\{\Id, \sigma_i\}$  of $N\times N$ matrices, and similarly $\rho$ with $\Tr[\rho]=1$, as
\be{rho:G}
G = \frac{1}{N}\, (g_0\,\Id + \bfg\cdot\bfsigma), \quad \rho = \frac{1}{N}\, (\Id + \bfx\cdot\bfsigma),
\ee
then substituting into the defining relation $\dot{\rho} = G\,\rho + \rho\,G$ we obtain the  linear equations in $\{g_0, \bfg\}$
\bsub
\bea{g0:g:eqns}
\frac{N}{2}\,\bfxdot\cdot\bfsigma &=& (g_0 + \frac{2}{N}\,\bfx\cdot\bfg)\,\Id 
+  \big(g_0\,\bfx + \bfg + \sum_{ij} x_i\,g_j \{\sigma_i, \sigma_j\}\big)\cdot\bfsigma, \label{g0:g:eqns:a} \\
\Rightarrow (i)\;\;0&=&g_0 + \frac{2}{N} \bfx\cdot\bfg,\quad
(ii) \;\; \bfg + g_0\,\bfx + D\,\bfg = \frac{N}{2}\,\bfxdot, \;\;\; (D)_{kj}\defn \sum_i\,x_i\,d_{ikj}. \label{g0:g:eqns:b} 
\eea
\esub
Substituting $g_0$ from (i) into (ii) and defining $(X)_{kj}\defn -\tfrac{2}{N}\,x_k\,x_j$ we can solve for 
$\bfg$, and subsequently $g_0$ by substituting this latter result back into (i). The results lead to the expressions
\be{g0:g}
\big(\Id + X(\bfx) + D(\bfx)\big)\,\bfg = \frac{N}{2}\,\bfxdot, \quad g_0 = -\bfx\cdot\big(\Id + X(\bfx) + D(\bfx)\big)^{-1}\,\bfxdot. 
\ee  
Note that the matrix $\Id + X(\bfx) + D(\bfx)$ is invertible due to the presence of the  diagonal elements of $\Id$.
\bigskip

For the familiar case of $N=2$, the above equations reduce to (see Chapter 5.4 of \cite{Jamiolkowski:2004})
\be{g0:g:N2}
g_0 =\frac{1}{2}\,\frac{\bfx\cdot\bfxdot}{1-x^2} = 
\frac{1}{4} \frac{d}{dt}\left(\ln \Delta^{-1/2}\right), \quad
\bfg = \frac{1}{2}\,\bfxdot - g_0\,\bfx = \frac{1}{2} \Delta^{1/2}\,\frac{d}{dt}\left(\Delta^{-1/2}\,\bfx \right),
\quad \Delta=\frac{1}{4}\,(1-x^2)=\det(\rho).
\ee

\subsection{Linear equations for $G$ in terms of $\rho$ for arbitrary $N$: unitary evolution}
For the case of unitary evolution $\rho(t) = U(t)\,\rho(0)\,U^\dag(t)$ the eigenvalues of $\rho(t)$ remain constant. Specializing to $N=2$, we see from \Eq{g0:g:N2} above that $\Delta(t)=$ const. implying that $g_0=0$ and $G=\half\,\bfxdot(t)\cdot\bfsigma$, which readily produces 
$\dot{\rho}=G$ from \Eq{GQS:9.42}, using the fact $|\bfx(t)|^2=$ const. so that 
$\bfx(t)\cdot\bfxdot(t)=0$, which is essentially the HLC.
Further, since one can also expand $A(t)=\frac{1}{2}\,(a_0(t)\,\Id + \bfa(t)\cdot\bfsigma)$, substituting into  $\dot{A} = G\,A$ one easily derives
\be{A:G}
\dot{a}_0 = \half\,\bfxdot\cdot\bfa, \quad \bfadot = \half\,(a_0\,\bfxdot + i\,\bfxdot\times\bfa), \quad 
\frac{d}{dt}\det(A) = 0,\quad \det(A) = a_0^2 - \bfa^2. 
\ee
Thus, while $\det(\rho)$ remains constant along the curve $\rho\in\MN$, we have concurrently that 
$\det(A)$ remains constant along its horizontal lift in $\E$.
\bigskip

Again, since $G$ relies on $\bfxdot$ (and in $\bfx$ for general, non-unitary evolution), one must 
\tit{a priori} have an explicit expression for the geodesic $\bfx(t)$, say from solving the geodesic (Euler-Lagrange) equations for the Bures metric $ds^2_B(\bfx, \bfxdot)$ (given after \Eq{Bures:metric}). However, for $N=2$, we are dealing with the 2-sphere Bloch ball $S^2$ where we can always orient the sphere such that geodesic occurs in the equatorial plane $\theta=\pi/2$ with Cartesian coordinate trajectory $\bfx(t) = (\cos(\omega t), \sin(\omega t), 0)$. We then see that 
$\dot{\rho}=G=\thalf\bfxdot = \omega\,(-\sin(\omega t), \cos(\omega t), 0)$ leads to the familiar evolution equation
$\bfxdot = \mathbf{\Omega} \times\bfx$ with the constant `magnetic field' $\mathbf{\Omega} = \omega\,\hat{\mathbf{z}}$.
\bigskip

For $N>2$ it is difficult to make such analogous analytic statements as in the case $N=2$ primarily due to the more involved nature of the $f_{ijk}$ and $d_{ijk}$ in the Lie algebra of $SU(N)$.
For unitary evolution, we have $G=i\,[\rho,Y]$ for Hermitian $Y=Y^\dag$ and Hamiltonian
$H = \rho\,Y + Y\,\rho$. Expanding $Y = \tfrac{1}{N}(y_0\,\Id + \bfy\cdot\bfsigma)$ one  obtains 
\bsub
\bea{G:Y}
g_0\Id + \bfg\cdot\bfsigma &=& \sum_{ijk} y_i\,x_j\,f_{ijk}\,\sigma_k \equiv (\tilde{D} \bfy)\cdot\bfsigma, \quad (\tilde{D})_{kj} = \sum_i x_i\,f_{ijk}, \label{G:Y:a} \\
\Rightarrow (i)\;\; g_0 =0; &{}& (ii)\;\; \bfg = \tilde{D}\,\bfy. \label{G:Y:b}
\eea
\esub
Recall that $g_0 = -\frac{2}{N}\, \bfx\cdot\bfg$ \Eq{g0:g:eqns:b}, implying that $ \bfx\cdot\bfg=0$, which also follows directly from (ii) in \Eq{G:Y:b} due to the antisymmetry of the structure constants. Note, that  $\tilde{D}$ is not an invertible matrix, so we cannot solve for $\bfy$ in terms of $\bfg$. Also, $y_0$  is not determined by \Eq{G:Y:a} and thus can be chosen freely.
We can also form the Hamiltonian $H=\rho\,Y + Y\,\rho$  
to obtain
\bea{H:from:Y}
H' \defn \tfrac{N^2}{2} H &=& (y_0 + \frac{2}{N}\,\bfx\cdot\bfy)\,\Id 
+ (\bfy + y_0 \bfx + D\bfy)\cdot\bfsigma, \no
&\to& \left(\big[ \Id + X(\bfx) + D(\bfx) \big]\,\bfy\right)\cdot\bfsigma,\quad \trm{choosing}\;\; y_0=-\frac{2}{N}\,\bfx\cdot\bfy,
\eea
where the matrix appearing in \Eq{H:from:Y} is the same one appearing in \Eq{g0:g}  defining $\bfg$ in terms of  $\bfxdot$. Further, while $g_0=0$ 
implies $\bfx\cdot\bfxdot=0$ for  $N=2$, \Eq{g0:g} requires that for general $N$,  
$\bfx$ and $\big(\Id + X(\bfx) + D(\bfx)\big)^{-1}\bfxdot$ are orthogonal along the orbit $\rho(t)$.
\bigskip

To try to get some insight into the meaning of $\bfy$, we first note the radius of the sphere that pure states
live on. Employing the decomposition $\rho = \tfrac{1}{N}\,(\Id + \bfx\cdot\bfsigma)$ and invoking the condition $\rho^2=\rho$ yields $x^2 = \binom{N}{2} = \thalf N(N-1)$ and $D\bfx = (N-2)\bfx$, the latter of which is a condition on the $d_{ijk}$ (identically satisfied for $N=2$). From \Eq{H:from:Y} we can then write 
\bsub
\bea{B}
H' \defn \bfB\cdot\bfsigma \Rightarrow \bfB &=& \bfy - \frac{2}{N}(\bfy\cdot\bfx)\bfx + D\bfy, \no
&=& \bfy - \alpha(N-1)(\bfy\cdot\bfxhat)\bfxhat  + D\bfy, \quad \bfx = x\,\bfxhat,\quad x^2 \defn \alpha\thalf N(N-1), \label{B:line1} \\
\Rightarrow  \bfB_{\parallel} &\defn& (\bfB\cdot\bfxhat)\,\bfxhat \quad
\trm{with}\;\; 
 \bfB\cdot\bfxhat = (1-\alpha (N-1)\,(\bfy\cdot\bfxhat) + \bfxhat\cdot (D\bfy), \qquad \bfB=\bfB_{\parallel}+\bfB_{\perp},  \label{B:line2} 
\eea
\esub
where we have defined $0\le\alpha\le1$ as the fractional radius of the pure state radius $\binom{N}{2}$, and  $\bfB_{\parallel}$ and $\bfB_{\perp}$ are the components of the effective magnetic field $\bfB$ parallel and perpendicular to $\bfxhat$.
One implication of \Eq{B:line2} is that for $N=2$ ($D=0$) and pure states ($\alpha=1$) we have
$\bfB_{\parallel}=0$ and we can take 
$\bfB=\bfB_{\perp} \defn \bfy - (\bfy\cdot\bfxhat)\bfxhat \equiv B\,\bfzhat$, which is consistent with the equatorial geodesics governed by ($\theta=\pi/2$) and $\bfxdot = \bfB\times\bfx$. For $N>2$, $\bfB_\parallel \ne 0$ in general. However, $\bfB_\parallel$ generates rotations about $\bfxhat$, which essentially contributes a dynamical phase to the quantum state. The relevant portion for generating the geodesic evolution is $\bfB_{\perp} = \bfB-\bfB_\parallel$.

\section{The horizontal lift geodesic in $\HS=\E$ and the geometric mean operator}\label{sec:HL:geodesic}
In this section we will develop an explicit formula for the purification $A(s)\in\SN(\H)$ along the geodesic path, satisfying the HLC, between two mixed quantum states $\rho_1$ (initial) and $\rho_2$ (final) of unit trace, using the operator of their geometric mean 
$M \defn \rho_1^{-1/2}\,\sqrt{\rho_1^{1/2}\,\rho_2\,\rho_1^{1/2}\,} \,\rho_1^{-1/2}$.
First, we need a slight digression on the expression for geodesics on spheres \cite{Zyczkowski_2ndEd:2020}.

\subsection{Geodesics on Spheres}
In the previous sections we have been talking about the geodesic between two density matrices as geodesics in the bundle space $\E=\HS$.
But what exactly are these geodesics with respect to the Bures metric? and what can we say about them? The answer is
\begin{itemize}
\item A \tit{geodesic with respect to the Bures metric} is a projection of a \tit{geodesic in the unit sphere $S^{2 N^2-1}\subset\HS=\E$} embedded in the bundle space with the \tit{added condition} that the \tit{geodesic is horizontal}, i.e. \tit{orthogonal to the fibers $\F$} in $\E$.
\end{itemize}

To explore this, we need some facts about geodesics on spheres. 
We know that geodesics on a sphere, in any dimensions, are great circles, and it is instructive to first see these arise formally on $S^2$, as an illustrative example. We can then generalize this to geodesics on odd dimensional spheres with either real coordinates ($S^{2N+1}$), or complex coordinates 
($\C P^N = S^{2N+1}/S^1$).
The statement that seems intuitive, but requires justification, is that in real coordinates $X=(X^0,X^1,\ldots,X^N)$ the geodesic on a sphere $S^N\in\R^{N+1}$ parametrized by the affine parameter $\s$ is given by
\bsub
\bea{GQS:3.12}
\trm{unit sphere:\;} S^N\;  &\Rightarrow& X(\s)\cdot X(\s) = 1, 	\label{GQS:3.12:line1} \\
\trm{geodesic\;}\Rightarrow X(\s) &=& X(0) \, \cos(\s) + \dX(0)\,\sin(\s), \label{GQS:3.12:line2} \\
\trm{with}\;\; X(0)\cdot X(0) &=&  \dX(0)\cdot\dX(0)=1,\; \trm{and} \; X(0)\cdot\dX(0)=0.  \label{GQS:3.12:line3}
\eea
\esub
First, the equations for the geodesics can be derived from the Euler-Lagrange equations, with constrained Lagrangian
\be{GQS:3.11}
\L= \dhalf \dX\cdot\dX + \dhalf\Lambda (1- X\cdot X),
\ee
with $\Lambda$ a Lagrange multiplier, to ensure the constraint $X\cdot X=1$ \Eq{GQS:3.12:line1} defining the sphere $S^N$.
Varying \Eq{GQS:3.11} yields the equation
\bea{}
\dfrac{d}{d\s}\left( \dfrac{\pd\L}{\pd\dX} \right) - \dfrac{\pd\L}{\pd X} = \ddot{X} + X=0, \qquad \trm{setting\;} \Lambda=1,
\eea
whose general solution is given by \Eq{GQS:3.12:line2}. 
A simple calculation reveals that the solutions (geodesics) remain on the sphere
\bea{}
1&=&X(\s)\cdot X(\s), \no
&=& X(0)\cdot X(0)\,\cos^2(\s) + \dX(0)\cdot\dX(0)\, \sin^2(\s) + 2\,X(0)\cdot\dX(0)\,\sin(\s)\,\cos(\s),
\eea
 for all values of $\s$ if the conditions  \Eq{GQS:3.12:line3} hold.
In particular, note that the tangent vector $\dX(0)$ is \tit{orthogonal} to the initial vector $X(0)$ defining the starting point of the geodesic.
The  tangent vector  $\dX(\s)$, initially orthogonal to $X(0)$, will in general rotate as it evolves along the evolved curve $X(\s)$, \tit{unless}  the curve is a geodesic, for which   $X(\s)\cdot\dX(\s) =0 $ remains valid all along the curve. In fact, this \tit{is} the definition of a geodesic, i.e. that the initial tangent vector to the curve, remains a tangent vector along the curve. Equivalently, the tangent remains \tit{parallel} to the curve. This is exactly the HLC discussed above.
First, let's prove that  $X(\s)\cdot\dX(\s) =0$ for a geodesic. In a calculation similar to the one above we easily have
\bea{}
1&=&X(\s)\cdot \dX(\s), \no
&=& \left( X(0) \, \cos(\s) + \dX(0)\,\sin(\s) \right) \,  \left( -X(0) \, \sin(\s) + \dX(0)\,\cos\s) \right), \no
&=& X(0)\cdot \dX(0)\,\left( \cos^2(\s) - \sin^2(\s) \right)  
+ \left( \dX(0)\cdot\dX(0) - X(0)\cdot X(0) \right)\,\sin(\s)\,\cos(\s), \qquad \no
&=& 0,
\eea
when we impose  \Eq{GQS:3.12:line3} .
The parameter $\s$ measures the (Bures) distance (arc length) $d_B= |\s_2-\s_1|$ along the geodesic, as can be seen by a short calculation.
Let $X_1= X(\s_1)$ and $X_2= X(\s_2)$ be any two points along the geodesic. Then we have
\bea{GQS:387}
X_1(\s_1) \cdot X_2(\s_2) &=&  \left( X(0) \, \cos(\s_1) + \dX(0)\,\sin(\s_1) \right) \,  \left( X(0) \, \cos(\s_2) + \dX(0)\,\sin(\s_2) \right), \qquad \no
&=& \cos(\s_1)\, \cos(\s_2) + \sin(\s_1)\, \sin(\s_2), \no
&=& \cos(\s_2-\s_1) \equiv \cos d_A. 
\eea

The extension of \Eq{GQS:3.12} to complex coordinates $Z = (Z^0,Z^1,\ldots,Z^N)$ is given by 
\bsub
\bea{GQS:3.85}
\trm{unit sphere:\;} S^{2N+1}\;  &\Rightarrow& Z(\s)\cdot \bZ(\s) = 1, 	\label{GQS:3.85:line1} \\
\trm{geodesic\;}\Rightarrow Z(\s) &=& Z(0) \, \cos(\s) + \dZ(0)\,\sin(\s), \label{GQS:3.85:line2} \\
\trm{with}\;\; Z(0)\cdot \Zbar(0) &=& \dZ(0)\cdot \dZbar(0)=1,\; \trm{and} \; Z(0)\cdot\dZbar(0) + \dZ(0)\cdot\Zbar(0)=0.  
\label{GQS:3.85:line3}
\eea
\esub
Recall the space of pure states $\ket{\psi} = \sum_{\a=0}^N Z^\a \ket{e_\a}$ such that $\braket{\psi|\psi}=1 = Z\cdot\Zbar$ for the complex projective space $\C P^N = S^{2N+1}/S^1$ whose natural metric is the Fubini-Study (FS) metric, discussed earlier, (see Section 3.5 \cite{Zyczkowski_2ndEd:2020}). Again, the parameter $\s$ measures the \tit{(Bures) distance $d_B= |\s_2-\s_1|$ along the geodesic} now given by
\be{GQS:3.88}
\cos d_A \equiv \cos(\s_2-\s_1)  = \dhalf\,\left( Z_1\cdot \Zbar_2 + Z_2 \cdot \Zbar_1\right) = \re{Z_1\cdot \Zbar_2}.
\ee

Returning to our  fiber bundle, the \tit{horizontal lifted geodesic} $A(s)\in\HS=\E$ in the bundle, such that $\rho(s) = A(s)\,A^\dag(s)\in\P=\M$ in the base space, is given by (where we have henceforth changed parameterization from $\s\to s$)
\bsub
\bea{GQS:9.5:9.56}
A(s) &=& A(0)\,\cos(s) + \dA(0)\,\sin(s), \label{GQS:9.5:9.56:line1} \\
\hspace*{-5.in}  \tr{A(0) A^\dag(0)} &=& \tr{\dA(0) \dA^\dag(0)} =1, \quad \tr{A(0)\dA^\dag(0) + \dA(0) A^\dag(0) } = 0, \label{GQS:9.5:9.56:line2}  \\
\dA^\dag(0) A(0) &=& A^\dag(0) \dA(0) = \left(  \dA^\dag(0) A(0) \right)^\dag \; (\trm{HLC})
 \Leftrightarrow \dA(s) = G(s) A(s). \qquad\;\; \label{GQS:9.5:9.56:line3} 
\eea 
\esub
 \Eq{GQS:9.5:9.56:line1} and  \Eq{GQS:9.5:9.56:line2} are the statements that $A(s)$ is a geodesic on the sphere $S^{2N^2-1}\subset\HS=\E$. 
 The crucial \tit{additional} feature is the \tit{horizontal lift condition}  \Eq{GQS:9.5:9.56:line3}
which ensures that $A(s)$ is the   \tit{unique, horizontal lift} (geodesic) of a geodesic $\rho(s) = A(s) A^\dag(s)$ in the base space connecting $\rho_1$ and $\rho_2$. The horizontal lift condition ensures  that  $A(s)\in\HS=\E$ lie along a curve in the bundle space that is 
 \tit{orthogonal to the fibers} $\F$ above $\rho(s)$.

\subsection{The HLC for the geodesic between two density matrices $\rho_1$ and $\rho_2$}
Let us again consider two density matrices $\rho_1 = \rho(0)$ and $\rho_2 = \rho(s)$, with their purifications $A_1(0)\defn 
A(0)$ and $A_2(s)\defn A(s)$ in the fibers lying above them, connected by the horizontal lift geodesic $A(s)$ in  \Eq{GQS:9.5:9.56:line1} above.
Note that the \tit{horizontal lift condition} (HLC) $\dA = G A$ \Eq{GQS:9.5:9.56:line3} states that 
$A^\dag_1(0) A_2(0) = A^\dag(0) A(0)$ is \tit{Hermitian}. (Recall, $\rho(0) = A(0) A^\dag(0)$ has the $A-$operators ``the other way around").
Further,  $A_1(0)^\dag A_2(s) = A^\dag(0) A(s)$ remains Hermitian all along the horizontal lift for all $s$ via
\bea{}
A^\dag(0) A(s) &=& A^\dag(0) \left( A(0)\,\cos(s) + \dA(0)\,\sin(s) \right), \no
&=&  A^\dag(0) A(0)\,\cos(s) + A^\dag(0) \dA(0)\,\sin(s), \no
&=&  A^\dag(0) A(0)\,\cos(s) + \left( A^\dag(0) \dA(0) \right)^\dag\,\sin(s), \quad (\trm{by HLC}) \no
&=&  A^\dag(0) A(0)\,\cos(s) +  \dA^\dag(0) A(0)\,\sin(s),\no
&=& A^\dag(s) A(0), \no
&=& \left( A^\dag(0) A(s) \right)^\dag
\eea
This tells us something about the operator the transforms $A(0)$ into $A(s)$ along the geodesic horizontal lift. 
From the HLC $\dA = G A$, we know from \Eq{A:formal:soln:line3}  the operator is given by 
\be{PMA:Fidelity:1}
\dA(s) = G(s) A(s) \quad \Rightarrow \quad A(s) = \T \big[ e^{\int_0^s G(s') ds'} \big]\,A(0) \equiv M(s)\,A(0),  \;\; M^\dag(s) = M(s),
\ee
where $\T$ is the \tit{path ordering} operator. Note that the final equality $M^\dag(s) = M(s)$ follows from the Hermiticity of $G^\dag=G$.

Let us substitute the horizontal lift condition 
\Eq{PMA:Fidelity:1}
into the geodesic solution \Eq{GQS:9.5:9.56:line1}, to obtain
\bsub
\bea{PMA:Fidelity:2}
A(s) 	&=& A(0)\,\cos(s) + \dA(0)\,\sin(s), \no
	&=&  A(0)\,\cos(s) + G(0)\,A(0)\,\sin(s), \no
	&=& \big( \Id\,\cos(s) + G(0)\,\sin(s) \big)\,A(0), \label{PMA:Fidelity:2:line1} \\
	& \equiv & M(s)\,A(0), \qquad M^\dag(s) = M(s), \label{PMA:Fidelity:2:line2}
\eea
\esub
where $M$ is an explicit representation of the path ordered expression in \Eq{PMA:Fidelity:1}, which only depends on the operator $G(0)$.
Again we see that $M^\dag(s) = M(s)$ follows from the fact that $G^\dag(0) = G(0)$.

Now taking the derivative of \Eq{PMA:Fidelity:2:line2}, and again using the HLC we have
\bsub
\bea{PMA:Fidelity:3}
\dA(s) &=& G(s) A(s), \no
\Rightarrow  \dot{M}(s) A(0) &=& G(s) M(s) A(0), \no
\Rightarrow  \dot{M}(s)  &=& G(s) M(s), \quad M(0) = \Id, \quad (\trm{since $A(0)$ is arbitrary}),
\label{PMA:Fidelity:3:line1}\\
\Rightarrow  M(s) &=& \T \big[ e^{\int_0^s G(s') ds'} \big] = \big( \Id\,\cos(s) + G(0)\,\sin(s) \big),\label{PMA:Fidelity:3:line2}
\eea
\esub
whose solution is consistent with \Eq{PMA:Fidelity:1} and \Eq{PMA:Fidelity:2:line1}.
\bigskip

Now using  $A(s) = M(s) A(0)$ \Eq{PMA:Fidelity:2:line2}, we easily see that
\bea{PMA:Fidelity:4}
A(s) = M(s) A(0) \; \Rightarrow \;\rho(s) &=& A(s) A^\dag(s) \no
&=& M(s) A(0) A^\dag(0) M(s), \no
&=& M(s) \rho(0) M(s). 
\eea
Thus, the origin of the initially cryptic looking statement $\rho(s) = M(s)\,\rho(0)\,M(s)$ is  simply a statement about the HLC condition between the respective purifications $A(0)$ and $A(s)$ of the initial and final states
for all $s$ along the horizontally lifted geodesic in the bundle of  the geodesic connecting $\rho(s)$ and $\rho(0)$ in the base manifold. 
\bigskip

Since $M$ is a positive Hermitian operator it can be interpreted as part of a dichotomous POVM 
$\{M(s), \Id-M(s)\}$. Therefore, the statement $\rho(s) = M(s)\,\rho(0)\,M(s)$ can be interpreted as the measurement $M(s)$ on $\rho(0)$ that yields the outcome $\rho(s)$ with unit probability since
$\Tr[M(s)\,\rho(0)\,M(s)] = \Tr[\rho(s)]=1$. In fact, it can be shown 
(see Chapter 9.4 of Wilde \cite{Wilde:2017}) that the eigenstates of $M(s)$ form the optimal basis from which to make measurements to maximize the root fidelity between $\rho(0)$ and $\rho(s)$.
\bigskip

We can solve for $M$ as follows. Let $\rho_1 = \rho(0)$ and $\rho_2 = \rho(s)$. 
Then, assuming  $\rho_1$ is invertible (or working only in its support), we have
\bea{PMA:Fidelity:5}
\rho_2 &=& M \rho_1 M, \no
\Rightarrow \tauopr &=& (\rhoonehalf M \rhoonehalf) \,  (\rhoonehalf M \rhoonehalf), \no
&=& (\rhoonehalf M \rhoonehalf)^2, \no
\Rightarrow \roottauopr &=&  (\rhoonehalf M \rhoonehalf) , \no
\Rightarrow  \rhooneinvhalf\, \roottauopr\,  \rhooneinvhalf &=&  M,
\eea
since $M>0$ has a unique square root.
As given by \Eq{PMA:Fidelity:5}, $M$  is called the \tit{geometric mean} $\rho_2 \# \rho_1^{-1}$ 
between $\rho_2$ and $\rho_1^{-1}$ \cite{Bhatia:2009,Zyczkowski_2ndEd:2020}.
\bigskip

In summary, for the 
Hilbert-Schmidt bundle space $\HS=\E$,  with positive (operator) cone base manifold $\P=\MN$, we have
\bsub
\bea{PMA:Fidelity:6}
A(s) &=& M (s)\,A(0)  \in\HS=\E, \quad \Rightarrow \rho(s) = M(s) \rho(0) M(s)  \in\P=\MN, \label{PMA:Fidelity:6:line1}  \\
\dot{M}(s) &=& G(s)\,M(s), \; M(0)=\Id,  \quad\; \rho(s) = A(s) A^\dag(s), \label{PMA:Fidelity:6:line2} \\
M(s) &=& \Id\,\cos(s) + G(0)\,\sin(s) =  \T \big[ e^{\int_0^s G(s') ds'} \big], \label{PMA:Fidelity:6:line3} \\
\Rightarrow M(s) &=& \Id\, \sqrt{F(s)} + G(0) \sqrt{1-F(s)}, \quad \cos(s) = \sqrt{F(s)}= \tr{M(s) \rho(0)} = \Tr\left[\sqrt{\rho^{1/2}(0)\rho(s)\rho^{1/2}(0)}\,\right],
\qquad\;  \label{PMA:Fidelity:6:line4} \\
 \trm{and}\quad M(s) &=& \rho^{-1/2}(0) \sqrt{\rho^{1/2}(0)\,\rho(s)\,\rho^{1/2}(0)} \,  \rho^{-1/2}(0) \ge 0. \label{PMA:Fidelity:6:line5}
\eea
\esub
The second to the last equality \Eq{PMA:Fidelity:6:line4} arises from the previous observation that the root fidelity $\rootF$ between two
arbitrary density matrices $\rho_1$ and $\rho_2$ is just the Bures angle $\cos D_A(\rho_1,\rho_2)$, between them, for which $D_A$ is just the difference in the affine parameters $D_A = |s_2-s_1|$ in the bundle along the horizontal lift (which exists on the unit sphere $S^{2 N^2-1}$ subspace in the bundle) of the geodesic connecting the two density matrices. 
It also follows by direct substitution of $M$ from \Eq{PMA:Fidelity:5} into the expression $\tr{M(s) \rho(0)}$ and observing that the cyclic property of the trace annihilates all the root powers of 
$\rho(0)$ \tit{outside} the square root - and directly gives the definition of the root fidelity.
The last equality follows from \Eq{PMA:Fidelity:5} and states that $M\ge0$ is a positive operator since $\rho(0)$ and $\rho(s)$ are both positive.

\subsection{Explicit horizontal geodesic between two purifications $A_1$ and $A_2$ lying above $\rho_1$ and $\rho_2$}
So far we have formally written the geodesic, satisfying the horizontal lift condition (HLC) in \Eq{PMA:Fidelity:6} in terms of the unspecified operator $G_0\equiv G(0)$, arising from the HLC $\dA^\dag(0) A(0) = \left( \dA^\dag(0) A(0) \right)^\dag = A^\dag(0) \dA(0)$, which is identically satisfied by the equation $\dA(s) = G(s) A(s)$ for $G^\dag = G$ Hermitian. We further showed that the geometric mean operator $M$ \Eq{PMA:Fidelity:6:line5} also satisfies the equation $\dot{M}(s) = G(s) M(s)$ with $M(0) = \Id$, so that $M(s)$ depends on $G_0$ as in \Eq{PMA:Fidelity:6:line3}. Lastly, we have $\rho(s) = A(s) A^\dag(s)\in\MN$ as the projection from the purification $A(s)\in\F\in\HS=\E$ in the fiber $\F$ in the bundle space $\HS=\E$ that sits above $\rho(s)$.
\bigskip

In this section, we want to explicitly display the form of the horizontal geodesic $A(s)$ in terms of the operator $M(s)$, given the two endpoint density matrices which we will again specify as $\rho_1 =\rho(s=0)\equiv\rho_0$ and $\rho_2 = \rho(s=\sstar) \equiv \rhostar$ associated with
$A(0)\equiv A_0$ and $A(\sstar)\equiv\Astar$
Here $\sstar$ is defined by the root fidelity between the initial and final states at the endpoints of the geodesic: $\cos(\sstar) = \rootF = \Tr\left[\sqrt{\rho^{1/2}\rho_2\rho_1^{1/2}}\,\right]$.
From the defining equation for the horizontal geodesic \Eq{PMA:Fidelity:6:line4} we have
\bsub
\bea{A:M:star}
A(s) &=& \big[\Id \cos(s) + G_0 \sin(s) \big] A(0) \defn M(s) A(0), \label{A:M:star:line1} \\
\trm{at}\; s=s^{*},\;  A(s)\to  A(\sstar) &=& \big( \Id \cosstar + G_0 \sinstar \big) A(0) \defn M(\sstar) A(0), \no
\Rightarrow G_0 &\defn& \dfrac{\Mstar - \Id \cosstar}{\sinstar}, \quad \trm{(since $A(0)$ is arbitrary)}
 \label{A:M:star:line2}\\
\Mstar \defn M(\sstar) &=&  \Mstaropr, \quad \rho_0 = A_0 A^\dag_0, \;\; \rhostar = \Astar \Astar^\dag = \rho(\sstar),  \label{A:M:star:line3} \\
\Rightarrow A(s) &=&  \left[\Id \cos(s) + \left( \dfrac{\Mstar - \Id \cosstar}{\sinstar} \right) \sin(s) \right] A(0) \defn M(s) A(0), \label{A:M:star:line4}
\eea
\esub
where in \Eq{A:M:star:line2} we can form $G_0$ by  \tit{explicitly constructing}  $\Mstar$ in \Eq{A:M:star:line3} given the beginning $\rho_0$ and endpoint $\rhostar$ of the projection (onto the base manifold $\MN$) of  horizontal geodesic between their respective initial $A_0$ and final $\Astar$ endpoints in $\E$.  
The last line \Eq{A:M:star:line4} gives the horizontal geodesic for all $0\le s \le \sstar$ with $A(0)=A_0$ and $A(\sstar) = \Mstar A_0$, the latter of which is the bundle analogue of  base manifold relationship  $\rhostar = \Astar \Astar^\dag = \Mstar \rho_0 \Mstar$.
\smallskip

It is instructive to see \tit{explicitly} that \Eq{A:M:star:line4}  satisfies (i) all the conditions of the geodesic and (ii) that it also satisfies the HLC so that the geodesic threads its way through the bundle orthogonal to the fibers.
\smallskip

First, for the geodesic normalization condition we trivially have $\tr{A_0 A^\dag_0} = \tr{\rho_0} = 1$.
Note also that explicitly we have  
$\Astar \Astar^\dag = \Mstar A_0 A^\dag_0 \Mstar  = \Mstaropr\, \rho_0\, \Mstaropr = \rhostar$, so that
$\tr{\Astar \Astar^\dag} = 1$.

From \Eq{A:M:star:line4} 
and \Eq{GQS:9.5:9.56:line1}
we have $\dA_0 = \left( \dfrac{\Mstar - \Id \cosstar}{\sinstar} \right) A_0$. 
Therefore, the second geodesic initial condition stating that the  \tit{tangent vector be of unit length}, is given by 
\bsub
\bea{unit:Adot}
\tr{\dA_0 \dA^\dag_0} &=&  \tr{ \left( \dfrac{\Mstar - \Id \cosstar}{\sinstar} \right) \rho_0  \left( \dfrac{\Mstar - \Id \cosstar}{\sinstar} \right)}, \no
&=&\dfrac{1}{\sinsqrdstar} \tr{\Mstar\rho_0\Mstar + \rho_0 \cossqrdstar - \cosstar \big(\Mstar\rho_0 + \rho_0\Mstar \big) }, \no
&=&\dfrac{1}{\sinsqrdstar} \left[ \tr{\rho_s} + \tr{\rho_0} \cossqrdstar - 2\,\cosstar\,\tr{\Mstar\rho_0} \right],\\
&=&  \dfrac{1}{\sinsqrdstar}  \left[ 1+ \cossqrdstar - 2\,\cosstar\,\cosstar\right], \no
&=& \dfrac{1-\cossqrdstar }{\sinsqrdstar}, \no
&=& 1
\eea
\esub
where in \Eq{unit:Adot} we have used the definition of the fidelity
$\tr{\Mstar\rho_0}$ $= \tr{\Mstaropr \rho_0}$ $= \tr{\roottauoprstar} \equiv \rootFstar \equiv \cosstar$ as the cosine of the Bures distance $d_A$,  $ \sqrt{F}(\rhostar,\rho_0) \equiv \cos d_A(\rho_0,\rhostar)$. 

The third geodesic initial condition is that $A_0$ and $\dA_0$ be orthogonal:
\bea{A0:dA0:orthogonal}
\tr{ \dA_0 A^\dag_0} &=& \tr{  \left(\dfrac{\Mstar - \Id \cosstar}{\sinstar} \, A_0\right) A^\dag_0}, \no
&=&  \dfrac{\tr{\Mstar \rho_0} - \tr{\rho_0} \cosstar}{\sinstar}, \no
&=&  \dfrac{ \cosstar-  \cosstar }{\sinstar}, \no
&=& 0.
\eea

For the last remaining relationship left, we need to show that the horizontal lift condition holds, namely
\bsub
\bea{HLC}
\trm{HLC:}\;\; \dA^\dag_0 A_0 &=& A^\dag_0 \dA_0 \equiv \left(\dA^\dag_0 A_0\right)^\dag \quad\Rightarrow\quad \dA^\dag_0 A_0,\ \trm{is Hermitian}, \\
\Rightarrow \dA^\dag_0 A_0 &=&  \left(\dfrac{\Mstar - \Id \cosstar}{\sinstar} \, A_0\right)^\dag A_0, \no
&=&  A^\dag_0\, \left(\dfrac{\Mstar - \Id \cosstar}{\sinstar} \, \right)A_0f, \\
&\equiv& \left[ A^\dag_0\, \left(\dfrac{\Mstar - \Id \cosstar}{\sinstar} \, \right)A_0 \right]^\dag, \quad \trm{since}\;  \Mstar^\dag = \Mstar. \nonumber
\eea
\esub

Thus, we have explicitly shown that
\be{HLC:geodesic}
A(s) =  \left[\Id \cos(s) + \left( \dfrac{\Mstar - \Id \cosstar}{\sinstar} \right) \sin(s) \right] A(0) = M(s) A(0),
\ee
is the unique geodesic in the bundle space $\E$,  which additionally satisfies the HLC, that connects $A_0\inE$, with $\rho_0 = A_0 A^\dag_0\in\MN$ in the base space, and $\Astar\inE$, with $\rhostar = \Astar \Astar^\dag\in\MN$.
\bigskip

In summary, given $\rho_0=\rho(0)$ and $\rhostar=\rho(\sstar)$ as initial and final states with
$\cos(\sstar) = \sqrt{F}(\rho_0,\rhostar)$
we have
\bsub
\bea{HLC:geodesic:MaxMixed:to:pure:Summary}
M(s) &=&  \left[\Id \cos(s) + \left( \dfrac{\Mstar - \Id \cosstar}{\sinstar} \right) \sin(s) \right], \quad 
0\le s\le s^*  \no
&\equiv&   \left[f(s)\,\Id  + g(s)\,\Mstar \right], \label{HLC:geodesic:MaxMixed:to:pure:Summary:a} \\
  f(s) &\equiv&  \cos(s) \left( 1 - \dfrac{\tan(s)}{\tanstar} \right), \quad f(0)=1, \;  f(s^*) = 0, \label{HLC:geodesic:MaxMixed:to:pure:Summary:b} \\
 g(s) &\equiv& \dfrac{\sin(s)}{\sinstar} , \quad \hspace{0.85in} g(0)=0, \;  g(s^*) = 1, \label{HLC:geodesic:MaxMixed:to:pure:Summary:c} \\
M(0) &=& \Id, \quad \Mstar = M(s^*),    \quad \rootFstar = \cosstar = \Tr[\Mstar\,\rho(0)], \\
\rho(s) &=& M(s)\rho(0)M(s) = f^2(s)\rho(0) + g^2(s)\rho(\sstar) 
                        				     + f(s) g(s) \left(M_{\sstar} \rho(0) + \rho(0) M_{\sstar} \right), \label{HLC:geodesic:MaxMixed:to:pure:Summary:e} \\
\Tr[\rho(s)]  &=&  f^2(s) + g^2(s) + \cos(\sstar) f(s) g(s) = 1.	\label{HLC:geodesic:MaxMixed:to:pure:Summary:f}		     
\eea
\esub

Finally, as a consistency check, we have that 
$\sqrt{F}(\rho_0,\rhostar) = \Tr[M(s)\,\rho_0] = \Tr[f(s)\,\Id+ g(s)\,\Mstar)\,\rho_0]=$ 
$f(s) + g(s)\,\Tr[\Mstar\,\rho_0] = f(s) + g(s)\,\cos(\sstar) = \cos(s)$ 
(using  \Eq{HLC:geodesic:MaxMixed:to:pure:Summary:a}, \Eq{HLC:geodesic:MaxMixed:to:pure:Summary:b} and \Eq{HLC:geodesic:MaxMixed:to:pure:Summary:c})
for any point $s$ along the geodesic.
\bigskip

Note that for evolution between orthogonal states, $\sstar=\pi/2$ so that 
$\rootFstar = \Tr[M_{\sstar}\,\rho(0)] = \cosstar = 0$. The above then simplifies to 
\bsub
\bea{M:orthogonal:states}
 M(s) &=& \left[\cos(s)\,\Id  + \sin(s)\,\Mstar \right], \\
 \rho(s) = M(s)\rho(0)M(s) &=& \cos^2(s)\rho(0) + \sin^2(s)\rho(\sstar) 
                        				     + \cos(s) \sin(s) \left(M_{\sstar} \rho(0) + \rho(0) M_{\sstar} \right). 
\eea
\esub

\section{Illustrative Examples}\label{sec:Examples}
In this section we present three examples illustrating  formulas
\Eq{HLC:geodesic:MaxMixed:to:pure:Summary:a} - \Eq{HLC:geodesic:MaxMixed:to:pure:Summary:f}
 developed in
\Sec{sec:HL:geodesic}.
 The first example is the geodesic between the maximally mixed state and a pure state, in arbitrary dimensions - clearly, a non-unitary evolution. 
 The second example is the geodesic
between Werner states
$\rho(p) = (1-p)\,\Id/N + p\,\ket{\Psi}\bra{\Psi}$
with 
$\ket{\Psi} = \{\ket{GHZ}, \ket{W}\}$
in dimension $N=2^3$. 
In the limit $p\to1$ there are an infinite number of possible geodesics between the orthogonal pure states $\ket{GHZ}$ and $\ket{W}$, so we illustrate the results obtained with the geometric mean operator with other possible geodesics.
Lastly, we analytically computed the geodesic orbit between two given qubit mixed states 
within the Bloch sphere for dimension $N=2$.
\smallskip

\subsection{Horizontal geodesic between arbitrary pure state $\ket{\psi}\bra{\psi}$ and the maximally mixed state $\rho_{\star}=\Id/N$}
Let us now compute an illustrative example that can be carried out fully.
Let $\rho_0 = \rho_{\star}\equiv\Id/N$ be the maximally mixed state in dimension $N$, and let us compute its horizontal geodesic to an arbitrary pure state  
$\ket{\psi}\bra{\psi}$, using \Eq{HLC:geodesic}. First off we note that 
 $\cosstar = \rootFstar =   \tr{\roottauoprstar} = \tr{ \sqrt{\tfrac{\Id}{\sqrt{N}} \ket{\psi}\bra{\psi} \tfrac{\Id}{\sqrt{N}} }} 
 = \tfrac{1}{\sqrt{N}}\tr{ \sqrt{\ket{\psi}\bra{\psi}}} = \tfrac{1}{\sqrt{N}}\tr{ \ket{\psi}\bra{\psi}} = \tfrac{1}{\sqrt{N}}$, where we have used that
 $\left( \ket{\psi}\bra{\psi}  \right)^p =  \ket{\psi}\bra{\psi} $ for any power $p$ since the latter has eigenvalue unity.
 Hence, $\sinstar = \tfrac{\sqrt{N-1}}{\sqrt{N}}$ and $\tanstar = \sqrt{N-1}$. By a similar calculation we have explicitly
 $\Mstar = \Mstaropr = \sqrt{N}\,\Id \sqrt{\tfrac{\Id}{\sqrt{N}} \ket{\psi}\bra{\psi} \tfrac{\Id}{\sqrt{N}} } \sqrt{N}\,\Id = \sqrt{N} \,  \ket{\psi}\bra{\psi}$.
Therefore
\bsub
\bea{HLC:geodesic:MaxMixed:to:pure}
M(s) 
&=& \Id f(s) +  g(s) \, \sqrt{N} \, \ket{\psi}\bra{\psi}, \quad 0\le s\le s^* = \cos^{-1}(1/\sqrt{N}), \\
M(0) = \Id, \quad \Mstar &=& M(s^*) =  \sqrt{N} \,\ket{\psi}\bra{\psi}, \quad  
\rootFstar = \cosstar = \dfrac{1}{\sqrt{N}},\;\; \sinstar = \dfrac{\sqrt{N-1}}{\sqrt{N}},  \\
\sqrt{F(s)} &=& \Tr[M(s)\, \Id/N] = f(s) + \frac{1}{\sqrt{N}}\,g(s) =  \cos(s).
\eea
\esub
Then, 
\bsub
\bea{rho:s:HLC:geodesic:MaxMixed:to:pure}
\rho(s) &=& M(s) \rho_0 M(s) = \dfrac{1}{N} M^2(s), \no
&=& f^2(s) \, \dfrac{\Id}{N} + \left(g^2(s) + \dfrac{2}{\sqrt{N}}\,g(s)\,f(s)\right)  \,\ket{\psi}\bra{\psi}, \label{rho:s:HLC:geodesic:MaxMixed:to:pure:line1} \\
\rho(0) &=&  \dfrac{\Id}{N} \to \rhostar = \rho(s^*) =  \ket{\psi}\bra{\psi}, \label{rho:s:HLC:geodesic:MaxMixed:to:pure:line2}
\eea
\esub
where \Eq{rho:s:HLC:geodesic:MaxMixed:to:pure:line2} follows from
the properties of $f(s)$ and $g(s)$ \Eq{HLC:geodesic:MaxMixed:to:pure:Summary:b} and \Eq{HLC:geodesic:MaxMixed:to:pure:Summary:c}.
Clearly, \Eq{rho:s:HLC:geodesic:MaxMixed:to:pure:line1} is not a unitary evolution since we have a maximally mixed state transforming into a pure state \Eq{rho:s:HLC:geodesic:MaxMixed:to:pure:line2} (i.e. the norms of the state are \tit{not} preserved along the horizontally lifted geodesic). 
Another, formal, yet informative way to display the horizontal lift transport operator $M(s)$ is through using 
$\rootFstar = \cosstar = \tfrac{1}{\sqrt{N}}$ to give from \Eq{HLC:geodesic:MaxMixed:to:pure}
\be{M:s:fidelity}
M(s) = \left[ \cos(s) \left( 1 - \dfrac{\rootFstar}{\rootoneminusFstar} \,\tan(s) \right) \Id
+ \dfrac{\sin(s)}{\rootFstar\,\rootoneminusFstar}   \ket{\psi}\bra{\psi} \right]. 
\ee

\subsection{Horizontal geodesic between Werner states with $\GHZ$ and $\W$ pure state components for $N=2^3$}
In this section we compute the horizontal geodesic between Werner states
$\rho_{\Phi}(p) = (1-p)\,\Id/N + p\,\ket{\Phi}\bra{\Phi}$ for $\ket{\Phi} = \{\GHZ, \W\}$ for three qubits $N=2^3$ where $\GHZ = \tfrac{1}{\sqrt{2}} (\ket{000} + \ket{111})$ and $\W = \tfrac{1}{\sqrt{3}} (\ket{001} + \ket{010} + \ket{100})$.
 We note that both $\rho_{GHZ}$ and $\rho_{W}$ have the same spectrum with eigenvalues
 $\lambda_1 = (1+ 7\,p)/8$, and $\lambda_2 = (1-p)/8$ with 7-fold degeneracy, but with different eigenstates. Using a symbolic computational program (such as Mathematica, used here) one can use the spectral representation of the $\rho$ to form quantities such as the initial state $\rho_1 = \rho_{GHZ}$,  
  final state $\rho_2 = \rho_{W}$, $\tau\equiv \rho_1^{1/2}\,\rho_2\, \rho_1^{1/2}$, 
  $\sqrt{\tau}$, and finally $\Mstar = \rho_1^{-1/2}\,\sqrt{\tau}\,\rho_1^{-1/2}$.
 From the root fidelity between $\rho_{GHZ}$ and $\rho_{W}$ one obtains
 \bsub
 \bea{rootF:GHZ:W}
 \sqrt{F_{\sstar}}\big(\rho_{GHZ}(p), \rho_{W}(q)\big) &=&
 \frac{1}{8}
 \left(
 6 \sqrt{(1-p)\,(1-q)} + \sqrt{(1-p)\,(1+7\,q)} +  \sqrt{(1-q)\,(1+7\,p)} 
 \right), \label{rootF:GHZ:W:a} \\
 \sqrt{F_{\sstar}}\big(\rho_{GHZ}(p), \rho_{W}(p)\big) &=&
 \frac{1}{4}
 \left(
 3\,(1-p) +  \sqrt{(1-p)\,(1+7\,p)}
 \right). \label{rootF:GHZ:W:b}
 \eea
 \esub
 \begin{figure}[h]
\begin{center}
\includegraphics[width=3.5in,height=1.75in]{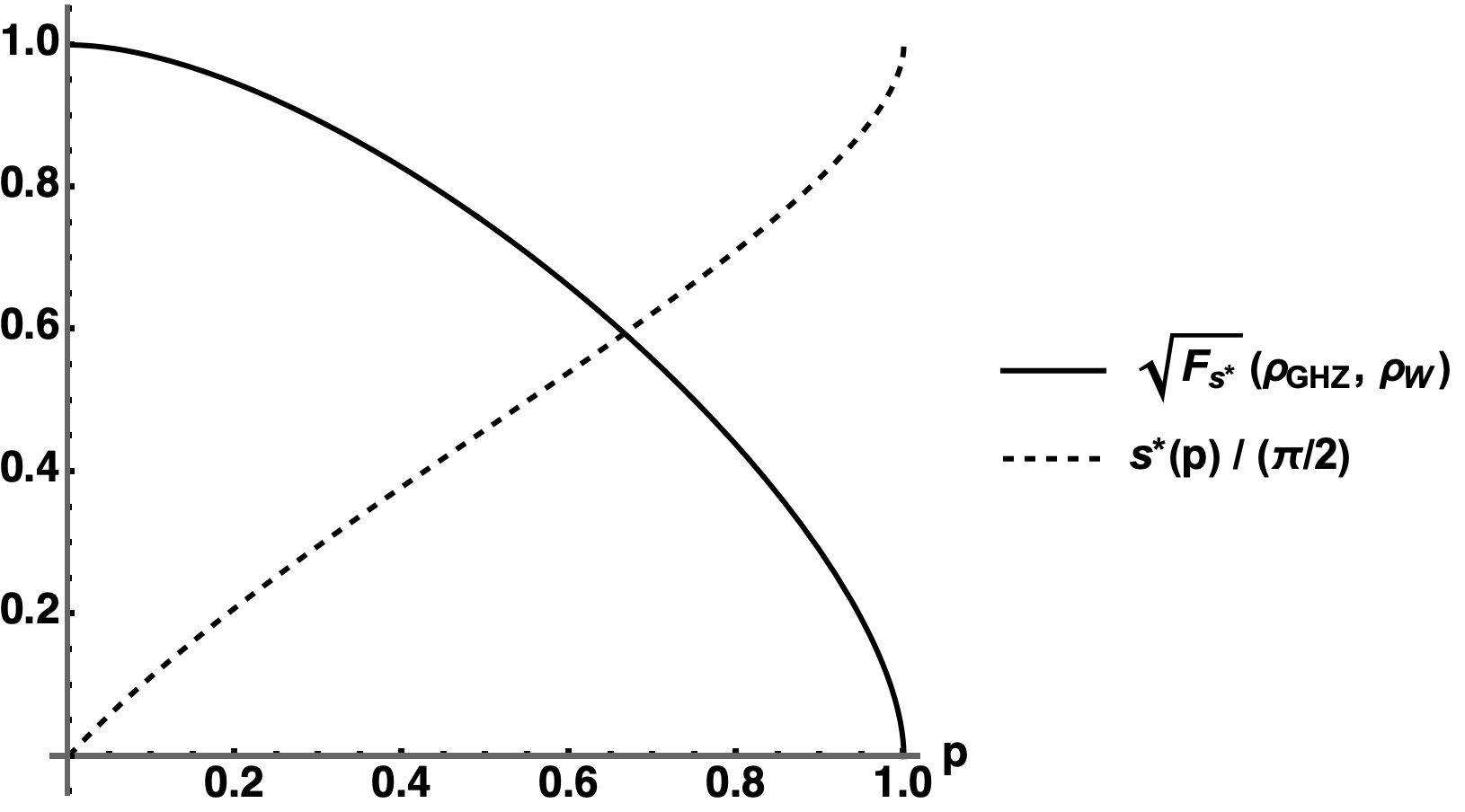}
\caption{Root Fidelity $\sqrt{F_{\sstar}}(p)=\cos\big(\sstar(p)\big)$ 
and Bures angle $\sstar(p)/(\pi/2)$ between the Werner states 
$\rho_{GHZ}(p)$ and $\rho_{W}(p)$ as function of the probability $p$. 
}\label{fig:rootF_sstar_rhoGHZ_rhoW}
\end{center}
\end{figure}
In \Fig{fig:rootF_sstar_rhoGHZ_rhoW} we plot 
$\sqrt{F_{\sstar}}\big(\rho_{GHZ}(p), \rho_{W}(p)\big) = \cos(\sstar)$ with equal probability $q=p$ for each state, and $\sstar = \sstar(p)$ for $0\le p \le 1$. At $p=0$ both states are the maximally mixed state $\Id/8$ so that $\sqrt{F_{\sstar}}=1$. At $p=1$, $\rho_{GHZ}\to\rhoGHZpure$ and $\rho_{W}\to\rhoWpure$ which are orthogonal $\sqrt{F_{\sstar}}=0$ with $\sstar=\pi/2$. 
\bigskip

Note that in \Eq{rootF:GHZ:W:b} 
 we are at fixed (final) arclength $s=\sstar$. Varying the Werner probability parameter $p$ varies the initial and final endpoint states along the geodesic $\rho(s)$. Thus, \Fig{fig:rootF_sstar_rhoGHZ_rhoW}
 shows the fidelity as a function of varying the endpoint states of the geodesics as one varies $p$.
For fixed $p$, that is, fixed endpoint states on the geodesic, the fidelity $\sqrt{F}(\rho_0,\rhostar)$ is always equal to $\cos(s)$ (i.e. the definition of the Bures geodesic distance), as shown in consistency check above after \Eq{HLC:geodesic:MaxMixed:to:pure:Summary:f}.
\bigskip

We can form the evolution operator $M(s)$ in \Eq{HLC:geodesic:MaxMixed:to:pure:Summary:a} and hence  $\rho(s)$ from \Eq{HLC:geodesic:MaxMixed:to:pure:Summary:e} to yield
\bsub
\bea{rho:s:rhoGHZ:rhoW}
\rho(s) &=& f^2(s)\,\rho_{GHZ} + g^2(s)\,\rho_{W} + f(s)\,g(s)\, \big(\Mstar\,\rho_{GHZ} + \rho_{GHZ}\,\Mstar \big), \label{rho:s:rhoGHZ:rhoW:line1}\\
 f(s) &=& \cos(s)\,\left(1- \dfrac{\tan(s)}{\tan(\sstar)}\right),  \quad g(s) = \dfrac{\sin(s)}{\sin(\sstar)},\quad
 \cos(\sstar) = \sqrt{F_{\sstar}} = \Tr[\Mstar\,\rho_{GHZ}],\\
 f(0)&=&1,\, f(\sstar)=0, \quad g(0)=0,\, g(\sstar)=1, \quad 
 \Tr[\rho(s)] = f^2(s) + g^2(s) + \cos(\sstar)\,f(s)\,g(s) = 1.
 \label{rho:s:rhoGHZ:rhoW:line2}
\eea
\esub
The third cross-term in \Eq{rho:s:rhoGHZ:rhoW:line1} is explicitly given by
\bea{crossterm}
\hspace{-0.25in}
\Mstar\,\rho_{GHZ} + \rho_{GHZ}\,\Mstar &=&
\frac{1}{4}\, \big( (1-p) +  \sqrt{(1-p)\,(1+7\,q)}\big)\,\rhoGHZpure 
-\frac{1}{4}\, (1-p)\, \big(\ket{000}\bra{111} + \ket{111}\bra{000} \big) \no 
-\frac{1}{4}\, (1-p)\,\rhoWpure &+& 
\frac{1}{4}\, (1-p)\,
\big(
\ket{001}\bra{001} +\ket{010}\bra{010} +\ket{100}\bra{100} 
+\ket{011}\bra{011} +\ket{101}\bra{101} +\ket{110}\bra{110} 
\big),\quad
\eea
which vanishes at $p=1$ although $f(0)\,g(0) = f(\sstar)\,g(\sstar)=0$ negates this term at the endpoints of the trajectory.
The second line of \Eq{crossterm} contains terms that are orthogonal to $\rhoGHZpure$.
At $p=1$ we have $\rho_{GHZ} \to \rhoGHZpure$, $\rho_{W} \to \rhoWpure$ with $\sstar=\pi/2$ so 
$\sqrt{F_{\sstar}}=0 = \Tr[M_{\pi/2} \rhoGHZpure]$
\bsub
\bea{rhos:s:p:1}
\rho(s) &\overset{p\to 1}{\longrightarrow}& \cos^2(s)\,\rhoGHZpure + \sin^2(s)\,\rhoWpure = A(s)\,A^\dag(s), \label{rhos:s:p:1:a} \\ 
A(s) &=&  \cos(s)\,\rhoGHZpure + \sin(s)\,\rhoWpure, \qquad\quad A(0) = \rhoGHZpure, \label{rhos:s:p:1:b} \\
\dot{A}(s) &=&  -\sin(s)\,\rhoGHZpure + \cos(s)\,\rhoWpure, \quad \quad \dot{A}(0) = \rhoWpure. \label{rhos:s:p:1:c}
\eea
\esub
\Eq{rhos:s:p:1:b} and \Eq{rhos:s:p:1:c} satisfy the HLC  $\dot{A}^\dag(0)\,A(0) = A^\dag(0)\,\dot{A}(0)$
since $A^\dag(0)\,\dot{A}(0) = \rhoGHZpure W\rangle\langle W| = 0$. 
Since $A^\dag(0)\,\dot{A}(0)$ is the operator analog of $\IP{\psi}{\dot{\psi}}$, 
\Eq{rhos:s:p:1:b} and \Eq{rhos:s:p:1:c} indicate that the connection is zero.
\bigskip

Note, even though we can define $A(s)$ as in \Eq{rhos:s:p:1:a} for $p=1$, we cannot form 
$M_{\sstar\to\pi/2}$ since the operator $\Mstar(p)$ is given by
\bea{Mstar:p}
\hspace{-0.25in}
\Mstar &=& \left(1+\sqrt{\frac{1-p}{1+ 7\,p}}\right)\,\rhoGHZpure 
- \big(\ket{000}\bra{111} + \ket{111}\bra{000} \big) \no
\hspace{-0.25in}
&+& \left(-1+\sqrt{\frac{1+ 7\,p}{1-p}}\right)\,\rhoWpure 
+ \ket{001}\bra{001} +\ket{010}\bra{010} +\ket{100}\bra{100} 
+\ket{011}\bra{011} +\ket{101}\bra{101} +\ket{110}\bra{110} 
\big),\quad
\eea
which is singular at $p=1$. In general we need the operator $\Mstar$ to provide us with the root fidelity between the initial and final states, which for $p=1$ is zero for orthogonal states. 
In fact, in attempting to use the formal definition of $M(s)$ in \Eq{PMA:Fidelity:5} we have
$\Mstar \overset{p\to 1}{\longrightarrow} 
\rho_{GHZ}^{-1/2}\,\left(|\IP{GHZ}{W}|\;\rhoGHZpure\right)\,\rho_{GHZ}^{-1/2}$ which is indefinite since 
$|\IP{GHZ}{W}|=0$ and $\rho_{GHZ}^{-1/2}$ is formally infinite since $\rhoGHZpure$ is not invertible in the space of $8\times 8$ matrices. Even on the support of $\rhoGHZpure$  where 
$(\rhoGHZpure)^m = \rhoGHZpure$ for any power $m$ (for eigenvalue $1$), we still have
$M_{\pi/2}  = \Mstar \overset{p\to 1}{\longrightarrow} 0$ as an operator. 
\bigskip

However, for orthogonal states, which by definition always have Bures angle $\sstar = \pi/2$, there are an \tit{infinite} number of geodesics connecting them. As a canonical example, in $N=2$ the 
`north' and `south' pole states $\ket{0}\bra{0}$ and $\ket{1}\bra{1}$ are connected by any longitudinal great circle. If we relax the requirement that $A$ needs to be represented by an $N\times N$ matrix, and can be represented by a pure state-vector (as discussed earlier), then taking $A(0)=\ket{GHZ}$ and
$A(\pi/2)=\ket{W}$ (as $8\times 1$ matrices)
so that $\rho(0)=\rhoGHZpure$ and $\rho(\pi/2)=\rhoWpure$,
we can define
 \bsub
 \bea{rho:s:alternative}
 A(s) &=& \cos(s)\,\ket{GHZ} + \sin(s)\,\ket{W}, \quad A(0) = \ket{GHZ},\;\; A(\pi/2) = \ket{W} = \dot{A}(0), \;\;  
 \trm{HLC:}\; A^\dag(0)\,\dot{A}(0)=0,\label{rho:s:alternative:a} \\
  \hspace{-0.5in}
 \Rightarrow \rho(s) = A(s)\,A^\dag(s) &=& \cos^2(s)\rhoGHZpure + \sin^2(s)\rhoWpure  
 + \cos(s)\sin(s)\,(\ket{GHZ}\bra{W} + \ket{W}\bra{GHZ}), \qquad  \label{rho:s:alternative:b} \\
 &\equiv&  \cos^2(s)\rhoGHZpure + \sin^2(s)\rhoWpure  + 
 \cos(s)\sin(s)\,(M_{\pi/2}\rhoGHZpure + \rhoGHZpure\,M_{\pi/2}), \qquad\; \label{rho:s:alternative:c} \\
 \trm{with} \;\; A(s) &=& M(s)\,A(0), \qquad M(s) = \Id \cos(s) + M_{\pi/2}\,\sin(s), \label{rho:s:alternative:d} \\
 M_{\pi/2}&=& \ket{GHZ}\bra{W} + \ket{W}\bra{GHZ}, \quad 
 \sqrt{F}(\rho_0,\rhostar) =\Tr[M_{\pi/2}\,\rhoGHZpure\,]= \cos(\sstar)=0,\, \sstar = \pi/2.
 \eea
\esub
In this case we \tit{can} assign a non-zero $M_{\pi/2}$ operator given by the intuitive choice 
$M_{\pi/2} =\ket{GHZ}\bra{W} + \ket{W}\bra{GHZ}$ which has the desired properties 
(i) $A(\pi/2) = M_{\pi/2}\,A(0)$ \Eq{rho:s:alternative:d} i.e. $\ket{W}=M_{\pi/2}\,\ket{GHZ}$,
  as well as the cross-term property 
  (consistency requirement of \Eq{rho:s:alternative:b} and \Eq{rho:s:alternative:c})
 (ii) $M_{\pi/2}\rhoGHZpure + \rhoGHZpure\,M_{\pi/2} = M_{\pi/2}$.
 Therefore, both of the geodesics between the orthogonal states $\rhoGHZpure$ and $\rhoWpure$ presented above are self-consistent, given their different representation $A(s)\in\E$. 
 The pure state representations $A(s)$ given in 
 \Eq{rhos:s:p:1:b} and \Eq{rho:s:alternative:a} are two of the infinite number of geodesics in $\E$ that project down to $\MN$ and geodesically connect the orthogonal state $\ket{GHZ}$ and $\ket{W}$.

\subsection{Geodesic between two arbitrary qubit density matrices $\rho_1$ and $\rho_2$ with the Bloch ball for $N=2$}
As a final example that yields tractable analytical results, we consider the geodesic between two arbitrary qubit density matrices $\rho_1$ and $\rho_2$ within the Bloch ball for $N=2$ (see \cite{Sjoqvist:2020,Alsing_Cafaro:ComparingMetrics:2023} for the computation of $\bfx(t)$ from solving the geodesic equations for $ds^2_B$ directly). As usual, the case $N=2$ for qubits  is tractable since the manipulation of $2\times 2$ matrices is manageable, with repeated use of the Pauli matrices property
$(\mathbf{a}\cdot\bfsigma)\,(\mathbf{b}\cdot\bfsigma)= \mathbf{a}\cdot\mathbf{b} + i\, (\mathbf{a}\times\mathbf{b})\cdot\bfsigma$. 
\bigskip

Let us write $\rho_1 = \half\,(\Id + \bfx\cdot\bfsigma)$ and  $\rho_2 = \half\,(\Id + \bfy\cdot\bfsigma)$. It is straightforward to show, by directly multiplication, that 
\be{matrices} 
\rho^{1/2}_1 = \tfrac{1}{\sqrt{2}}\,(a_-\,\Id + a_+\,\bfxhat\cdot\bfsigma), \quad
\rho^{-1/2}_1 = \tfrac{1}{\sqrt{2\,\det\rho_1}}\,(-a_-\,\Id + a_+\,\bfxhat\cdot\bfsigma), \quad,
\trm{with}\quad a_{\pm} = \sqrt{\thalf \pm \sqrt{\det\rho_1}}, \quad \bfx \defn |\bfx|\,\bfxhat,
\ee
where $\det\rho_1 = \tfrac{1}{4} (1-|\bfx|^2)$, so that alternatively
$a_\pm = \tfrac{1}{\sqrt{2}}\, \sqrt{1\pm\sqrt{1-x^2}}$ with $x^2 \defn |\bfx|^2$. 
A lengthy, but straightforward calculation yields
\be{tau:0:tau:vec}
\hspace{-0.5in}
\tau = \rho^{1/2}_1\,\rho_2\,\rho^{1/2}_1 = \tau_0\,\Id + \bftau\cdot\bfsigma, \quad
\tau_0 = \tfrac{1}{4}\,\big(1+x\, (\bfy\cdot\bfxhat)\big), \quad
\bftau = \tfrac{1}{4}\,
\big[
\big(x+ (\bfy\cdot\bfxhat)\big) \,\bfxhat 
-\thalf\,(1-x^2)\,\bfy_{\perp}
\big], \quad 
\bfy_{\perp}\defn \bfy-(\bfy\cdot\bfxhat)\,\bfxhat,
\ee
where $\bfy_{\perp}$ is perpendicular to $\bfxhat$.
Thus, $|\bftau| = \tfrac{1}{4} \left[ (x+ y_\parallel)^2 + \tfrac{1}{4}(1-x^2)^2 |\bfy_\perp|^2\,\right]^{1/2}$,
where we have defined $\bfy_\parallel = (\bfy\cdot\bfxhat)\,\bfxhat = y_\parallel\,\bfxhat$.
We can diagonalize $\tau$ to obtain its eigenvalues $\Lambda_\pm$ 
and eigenvectors $\ket{\Lambda_\pm}$, and hence construct $\sqrt{\tau}$,
leading to 
\bsub
\bea{root:tau}
\sqrt{\tau} &=& \sqrt{\rho^{1/2}_1\,\rho_2\,\rho^{1/2}_1} =  
\sqrt{\Lambda_+}\,\ket{\Lambda_+}\bra{\Lambda_+} +  \sqrt{\Lambda_-}\,\ket{\Lambda_-}\bra{\Lambda_-}, \label{root:tau:a} \\
\Lambda_\pm &=& \tau_0 \pm |\bftau|, \quad
\ket{\Lambda_\pm} =
\frac{1}{\sqrt{2\,|\bftau|\,(|\bftau|\pm \tau_3)}}
\left(
\begin{array}{c}
\tau_3\pm  |\bftau|\\
\tau_1 + \tau_2\end{array}
\right), \quad
\IP{\Lambda_\pm} {\Lambda_\pm}=1,\quad 
\IP{\Lambda_\pm} {\Lambda_\mp}=0. \label{root:tau:b}
\eea
\esub
From \Eq{root:tau:a} we can write down the fidelity between the geodesic endpoint states $\rho_1$ and $\rho_2$ as
\be{fidelity:N:2}
\sqrt{F_{\sstar}}(\rho_1,\rho_2) = \Tr[\sqrt{\tau}]= \sqrt{\Lambda_+}+\sqrt{\Lambda_-}
=\sqrt{\tau_0 + |\bftau|}+\sqrt{\tau_0 - |\bftau|} \defn \cos(\sstar).
\ee

From the above we could compute $\Mstar = \rho^{-1/2}_1\,\sqrt{\tau}\,\rho^{-1/2}_1$ directly. 
However, from \Eq{HLC:geodesic:MaxMixed:to:pure:Summary:e}
$\rho(s)= f^2(s)\rho_1 + g^2(s)\rho_2 + f(s) g(s) \left(M_{\sstar} \rho_1 + \rho_1 M_{\sstar} \right)$. Therefore, for the evolution of the states along the geodesic  the relevant quantity to compute 
is the crossterm  $M_{\sstar} \rho_1 + (M_{\sstar} \rho_1)^\dag$, where
$M_{\sstar} \rho_1 = \rho_1^{-1/2}\,\sqrt{\tau}\,\rho_1^{1/2}$. Using the above quantities, another lengthy but straightforward computation reveals that
\bsub
\bea{CT}
M_{\sstar} \rho_1 + (M_{\sstar} \rho_1)^\dag  &=& 
2\, \sum_{i=\pm}\, \sqrt{\Lambda_i}\, \rho^{(i)}, \quad
\rho^{(i)} = \thalf\,\left( \Id + (\mathbf{w}^{(i)}_{\parallel} - 
\tfrac{1}{\sqrt{1-x^2}}\,\mathbf{w}^{(i)}_{\perp})\cdot\bfsigma\right), \\
\mathbf{w}^{(i)} &=& \EV{\Lambda_i}{\bfsigma}{\Lambda_i}\defn \mathbf{w}^{(i)}_{\parallel} + \mathbf{w}^{(i)}_{\perp}, \qquad
\mathbf{w}^{(i)}_{\parallel} =  (\mathbf{w}\cdot\bfxhat)\,\bfxhat, \qquad
\mathbf{w}^{(i)}_{\parallel} = \mathbf{w} - (\mathbf{w}\cdot\bfxhat)\,\bfxhat.
\eea
\esub
Thus, the expression for the states $\rho(s)$ along the geodesic between the endpoint states $\rho_1$ and $\rho_2$ within the Bloch ball for $N=2$ is given by
\be{rho:s:N:2}
\rho(s)= f^2(s)\,\rho_1 + g^2(s)\,\rho_2 
+ 2\, f(s)\,g(s)\, \sum_{i=\pm} \sqrt{\Lambda_i}\; \rho^{(i)}, \qquad 0\le s \le \sstar,
\ee
where, again, from \Eq{HLC:geodesic:MaxMixed:to:pure:Summary:b} and 
\Eq{HLC:geodesic:MaxMixed:to:pure:Summary:c}
we have $f(0)=1,\, g(0)=0$ and $f(\sstar)=0,\, g(\sstar)=1$, with 
$\sstar$ defined in \Eq{fidelity:N:2}. Defining $\rho(s)\defn \thalf\,(\Id + \mathbf{r}(s)\cdot\bfsigma)$ we can extract the geodesic orbit $\mathbf{r}(s)=\Tr[\rho(s)\,\bfsigma]\in\mathbb{R}^3$ within the Bloch ball from \Eq{rho:s:N:2} as
\be{r:s}
\mathbf{r}(s) = f^2(s)\,\bfx + g^2(s)\,\bfy 
+ 2\,f(s)\,g(s) \sum_{i=\pm} \sqrt{\Lambda_i}\,
\left(\mathbf{w}^{(i)}_{\parallel} - 
\tfrac{1}{\sqrt{1-x^2}}\,\mathbf{w}^{(i)}_{\perp}\right).
\ee

\section{Summary and Conclusion}\label{sec:Summary:Conclusion}
In this work we have explored consequences of  horizontal lift condition (HLC) which determines the geodesic in the Hilbert Schmidt bundle space that threads orthogonally to the fibers, and projects down to the geodesic in the base manifold between two mixed states. The HLC leads to the well-known non-unitary evolution equation $\dot{\rho} = G\,\rho + \rho\,G$ governed by the logarithmic derivative of $\rho$, for $G$ a Hermitian operator. We explored expressions for $G$ in terms of $\rho$ using an expansion of the operators in terms of the $N^2-1$ generators of $SU(N)$ and the identity matrix. In this representation 
$\rho$  depends on the Cartesian orbit $\bfx(t)\in\mathbb{R}^{N^2-1}$, and therefore in general $G= G\big(\bfx(t), \bfxdot(t)\big)$. Hence, one must \tit{a priori} obtain the path $\bfx(t)$ by other means, say by solving the (Euler-Lagrange) geodesic equations for the Bures metric $ds^2_B$. The expressions for the parameters defining $G$ in terms of $\bfx(t)$ and  $\bfxdot(t)$ describing $\rho$ are linear, 
as summarized by \Eq{g0:g}, and can in principle be readily solved in any dimension $N$. We explored these expressions for both non-unitary and unitary evolutions of $\rho$. 
\smallskip

Additionally, we presented an alternate means to describe the mixed states along a geodesic in $\MN$ connecting two fixed, initial and final, endpoint states $\rho_1$ and $\rho_2$, respectively,
as summarized in \Eq{HLC:geodesic:MaxMixed:to:pure:Summary:a}- \Eq{HLC:geodesic:MaxMixed:to:pure:Summary:f}. 
This was accomplished through the use of the geometric mean operator between these two end states, which allowed us to construct  purifications all along the horizontal lifted geodesic in $\E$. We demonstrated this approach on the non-unitary evolution between a maximally mixed state and a pure state in an arbitrary dimension $N$. We also presented the explicit (in general, non-unitary) geodesic between Werner states  for 3-qubits ($N=2^3=8$) with probability $p$ to have pure state components of $\{\ket{GHZ},\,\ket{W}\}$, and probability $1-p$ for the maximally mixed state. We examined the expression for the geodesic in both 
$\E$ and $\MN$ in the limit  $p \to1$ where  there are now an infinite number of possible geodesics between the orthogonal pure states.
Lastly, we analytically computed the geodesic orbit \Eq{r:s} between two given, arbitrary qubit mixed states 
within the Bloch sphere for dimension $N=2$.
\smallskip

It is well known that one cannot transform $\ket{GHZ}$ to $\ket{W}$ by means of local unitary - one requires non-local interactions \cite{Uskov_Alsing:2020,Uskov_Alsing:2023}. It would be interesting to compare the geodesics and fidelity obtained here between Werner states with $\{\ket{GHZ},\,\ket{W}\}$ pure state components to the actual efficiencies of such proposed transformation interactions. 
Further, much research has been conducted on quantifying the efficiency, by various measures, of the evolution between quantum states, both pure and mixed 
(see e.g. \cite{Cafaro_Alsing:2022a,Cafaro_Alsing:2022b} and references therein). Since the, in general non-unitary, geodesic between two states represents the optimal transition probability between the two mixed states, it would be interesting to determine how well such geodesics could be approximated by a single, or a series of unitary evolutions.  
\bigskip

As to the relevance of the work presented here, we have shown that the geometric mean operator formalism provides a very compact and useful recipe for finding the geodesic paths
connecting any two arbitrary full-rank $N\times N$ density matrices.
The presence of the Uhlmann fidelity between the initial and 
intermediate states along the geodesic is made manifest in this representation.
Importantly, the initial and final mixed states do not need to be
isospectral. This way, one can deal with both unitary and nonunitary quantum
geodesic evolutions. More specifically, we see two relatively
straightforward applications of our formalism. The first one concerns
extending the study on the complexity of geodesic evolutions for systems
beyond single-qubit pure states \cite{Cafaro_Alsing:2022a}. Despite its
apparent conceptual simplicity, this would constitute the achievement of a
nontrivial quantitative task \cite{brown19,chapman18,ruan21}. The second
one, instead, consists in quantifying in an explicit fashion the deviations
from the geodesicity property of (both unitary and nonunitary) quantum
evolutions beyond two-level (closed) quantum systems evolving unitarily \cite%
{Cafaro_Alsing:Qubit:Geodesics:2023}. This would be a highly valuable achievement given that (to
the best of our knowledge) there exist only algorithms for solving
unconstrained quantum brachistochrone problems in a unitary setting limited
to (minimally energy wasteful \cite{uzdin12}) paths connecting two
isospectral mixed quantum states via a fast evolution \cite{campaioli19}.

We wish our findings will motivate other scientists and pave the way toward
further investigations in this fascinating research direction on the use of
the geometry of quantum states to better understand quantum physics. For the
moment, we postpone a more quantitative discussion on these potential
extensions and applications of our analytical findings to future scientific
investigations.

\vspace{-0.25in}
\section{Acknowledgements}
 Any opinions, findings and conclusions or recommendations expressed in this material are those of the author(s) and do not necessarily reflect the views of their home institutions.
%
\vspace{-0.25in}
\appendix
\section{Proof that $U= \sqrt{\rho_1^{1/2}\,\rho_2\,\rho_1^{1/2}\,}\,\rho_1^{-1/2}\,\rho_2^{-1/2}$ is unitary}\label{app:A}\
In the derivation for the Uhlmann fidelity $\rootF = \Tr[\sqrt{\rho_1^{1/2}\,\rho_2\,\rho_1^{1/2}}]$ one is required to maximize the expression in \Eq{GQS:9.35}
\be{GQS:9.35:again}
\tr{A_1 A_2^\dag} = \tr{   \rootrhoone U_1\,  (\rootrhotwo U_2)^\dag\,} 
= \tr{ U\, \rootrhotwo  \, \rootrhoone \, }, \quad U =U_1 U_2^\dag .
\ee
After taking polar decompositions for the positive (invertible) matrices $A_1$ and $A_2$, one might \tit{almost} naively  guess from \Eq{GQS:9.35:again} that 
$U = \roottau \, \rhooneinvhalf \, \rhotwoinvhalf$ where we have defined $\tau = \tauopr$ so that 
$\rootF = \tr{\, U \rootrhotwo  \rootrhoone }$ 
$=  \tr{\, \roottau\, \rhooneinvhalf \rhotwoinvhalf \rootrhotwohalf \rootrhoonehalf }  =\tr{\roottau\,}$. 
The \tit{almost} qualifier comes from the fact that we must show that this choice of $U$ is indeed unitary.
We now show that this is indeed the case

Define $A \equiv \rootrhoone\,\rootrhotwo$ so that $A A^\dag = (\rootrhoone\,\rootrhotwo) (\rootrhoone\,\rootrhotwo)^\dag = 
\rootrhoone \rho_2 \rootrhoone \equiv \tau$. We note the following: $\Ainv =\rhotwoinvhalf \rhooneinvhalf $ and
$ \Adaginv= \rhooneinvhalf  \rhotwoinvhalf $. Thus, 
\bea{PMA:U}
U &=& \roottau \, \rhooneinvhalf \, \rhotwoinvhalf  = \sqrt{A A^\dag} \, \Adaginv, \quad A \equiv \rootrhoonehalf\,\rootrhotwohalf\\
%
%
\Rightarrow\; U \, U^\dag &=&  (\sqrt{A A^\dag} \, \Adaginv) \, (\Ainv \, \sqrt{A A^\dag}), \no
&=& \sqrt{A A^\dag}\, (A A^\dag)^{-1} \sqrt{A A^\dag}, \no
&\equiv& \Id, \\
%
%
\Rightarrow\; U^\dag\, U &=&  (\Ainv \, \sqrt{A A^\dag}) \,  (\sqrt{A A^\dag} \, \Adaginv), \no
&=& \Ainv\, (A A^\dag) \, \Adaginv, \no
&\equiv& \Id.
\eea
Thus, somewhat surprisingly, $U= \sqrt{A A^\dag} \, \Adaginv$ is unitary, if $A$ is invertible.

%

\end{document}